\documentclass[final]{ws-ijmpd-nocrop}
\usepackage{bezier}
\usepackage{graphicx}
\usepackage{fancyvrb}

\begin{document}

\newcommand{\sla}[1]{#1\!\!\!/}


\title{An Algorithmic Approach to Quantum Field Theory}

\author{Massimo Di Pierro}

\address{School of Computer Science, Telecommunications and Information Systems,\\
DePaul University, 243 S Wabash Ave, Chicago, IL 60604\\
mdipierro@cs.depaul.edu}

\maketitle


\begin{abstract}
The lattice formulation provides a way to regularize, define and compute the Path Integral in a Quantum Field Theory. In this paper we review the theoretical foundations and the most basic algorithms required to implement a typical lattice computation, including the Metropolis, the Gibbs sampling, the Minimal Residual, and the Stabilized Biconjugate inverters. The main emphasis is on gauge theories with fermions such as QCD. We also provide examples of typical results from lattice QCD computations for quantities of phenomenological interest.
\end{abstract}


\section{Introduction}

In this paper we present a brief review of Lattice Quantum Field Theory (LQFT), a way to formulate a Quantum Field Theory (QFT) in algorithmic terms%
\footnote{For an introduction to QFT see \cite{peskin,itzykson} and
for LQFT see \cite{creutz,rothe,montvay,gupta}.}.

Most of this work is based on lectures given at Fermilab by the author in 2001.

QFTs are the application of quantum mechanics to fields. They form a very general class of mathematical models that reduces to quantum mechanics in the non-relativistic limit (speed of light $\rightarrow\infty$), to relativistic mechanics in the decoherence limit (Plank constant $\rightarrow 0$) and to classical physics when both limits are taken.

Any QFT states that%
\footnote{From now on we assume units in which both the Plank constant and the speed of light are 1.}:
\begin{itemize}
\item
Each type of elementary particle ``A'' is associated with a field $\phi(x)$ so that $|\phi(x)|^2$ represents the probability of the event ``particle A is at the space-time point $x=(x_0,\mathbf{x})$'';

\item
There exists a functional of the field $\mathcal{S}[\phi,...]$, called action, which describes the dynamics of the system and any correlation amplitude between multiple time ordered events A at $x$, A at $x'$, A at $x''$, etc. and can be computed as
\begin{equation}
\left<0\right|\phi(x) \phi(x') \phi(x'') ...\left|0\right> \stackrel{def}{=}
\int [\textrm{d}\phi] \phi(x) \phi(x') \phi(x'') ... e^{i \mathcal{S}[\phi,...]}
\label{qft1}
\end{equation}
The square of a correlation amplitude is a regular real correlation function.

\end{itemize} 
The field components $\phi(x)$ can be scalars, complex, spinors, vectors, tensors, etc., depending on the properties of the particle A. $x'$, $x'$, $x''$ are points in the space-time. The symbol $\int [\textrm{d}\phi]$ indicates an integral over a Hilbert space and is known as Path Integral (PI). Each $\phi$ in the Hilbert space is referred to as a {\it history} or {\it path} or {\it field configuration}. 

The symbol $\phi$ is used here as a template for a generic field and, for now, one may think of it as a real scalar field. Later $\phi$ will be replaced by the symbol $U$ when it represents a gauge field and by the symbol $\psi$ when it represents a spinor, for example, a quark.

Translational invariance combined with the finite speed of light implies that $\mathcal{S}$ can be written as the integral over the $s$-dimensional space-time of a local function of $\phi$, the Lagrangian density $\mathcal{L}$.
\begin{equation}
\mathcal{S}[\phi,...]=\int \mathcal{L}(\phi(x),\partial_\mu\phi(x)...) \textrm{d}^s x
\label{qft2}
\end{equation}

The dots indicate additional fields and/or parameters of the theory such as particle masses and coupling constants. The Lagrangian is usually a function of $\phi(x)$, it's gradient and higher derivatives.

The naive assumption that the Hilbert space in the PI is the space of all possible distributions leads to the problem of divergences of QFTs. In order to understand and cure this problem one has to properly define this space and provide an algorithmic way to evaluate the integral. 
That is the main purpose of these notes.

For some theories, such as quantum electrodynamics (QED), it is possible to perform a functional expansion of the integrand of eq.~(\ref{qft1}) around a minimum and integrate exactly the individual terms. These terms are the Feynman diagrams. The result is an asymptotic series, the Dyson series, that can be used to approximate the PI. While this is a well defined algorithm it is very difficult to implement, it only works in some cases, and it cannot be improved to arbitrary accuracy because of the asymptotic nature of the Dyson series. The divergences that appear in this case can be subtracted by regulating the theory. For example, one can dump the Fourier modes of the fields with frequency above some cut-off $p$.

The perturbative expansion plays a role analogous to the Taylor expansion and it does not provide a general definition of the PI. Moreover, for some theories, the Dyson series may diverge and a different approach is required in order to define and compute eq.~(\ref{qft1}). 

For quantmm chromodynamics\cite{qcdreview,marciano} (QCD) and other strongly coupled theories, the lattice formulation has been one of the most successful. We will refer to the latter as lattice quantum chromodynamics (LQCD).

\section{Overview}

\subsection{Formulation}

The formulation of LQFT consists of three stages\cite{creutz,rothe,montvay}:

{\bf Step 1: Discretization.} The space-time is approximated with a finite hypercubic mesh such that $\phi(x)$ is only defined on the lattice sites $x$. Now the Hilbert space that constitutes the integration domain is well defined. The hypercubic mesh is characterized by the number of dimensions, $s$, the lattice spacing, $a$, the overall size, $L=Ka$, and the boundary conditions. Here is an example of a 3D lattice.
\begin{center}
\setlength{\unitlength}{.02in}
\begin{picture}(75,75)(0,0)
\multiput(25,25)(10,0){4}{\circle*{3}}
\multiput(25,35)(10,0){4}{\circle*{3}}
\multiput(25,45)(10,0){4}{\circle*{3}}
\multiput(25,55)(10,0){4}{\circle*{3}}

\multiput(28,27)(10,0){4}{\circle*{2}}
\multiput(28,37)(10,0){4}{\circle*{2}}
\multiput(28,47)(10,0){4}{\circle*{2}}
\multiput(28,57)(10,0){4}{\circle*{2}}

\multiput(31,29)(10,0){4}{\circle*{2}}
\multiput(31,39)(10,0){4}{\circle*{2}}
\multiput(31,49)(10,0){4}{\circle*{2}}
\multiput(31,59)(10,0){4}{\circle*{2}}

\put(66,44){\vector(0,1){15}}
\put(66,44){\vector(0,-1){15}}
\put(71,44){\makebox(0,0){$L$}}

\put(50,20){\vector(1,0){5}}
\put(50,20){\vector(-1,0){5}}
\put(50,15){\makebox(0,0)[t]{$a$}}

\put(11,55){\vector(1,0){8}}
\put(9,55){\makebox(0,0)[r]{$\phi(x)$ at site $x$}}

\end{picture}
\end{center}
After discretization, for every finite $a$ the symbol $\int [\textrm{d}\phi]$ becomes a well-defined multidimensional integral 
\begin{equation}
\int [\textrm{d}\phi] \simeq \int \textrm{d}\phi_0 \int \textrm{d}\phi_1 \int \textrm{d}\phi_2 ...
\label{qint}
\end{equation}
where the $i$-th degree of freedom, $\phi_i$ is localized at some lattice site.

{\bf Step 2: Wick rotation.} 
The time coordinate $x_0$ is Wick rotated $x_0 \rightarrow i x_0$ thus turning eq.~(\ref{qft1}) into
\begin{equation}
\left<0\right|\phi(x) \phi(x') \phi(x'') ...\left|0\right>_E =
\int [\textrm{d}\phi] \phi(x) \phi(x') \phi(x'') ... e^{-\mathcal{S}_E[\phi]}
\label{qft3}
\end{equation}
where $\mathcal{S}_E$ is now the Euclidean action.
If the exponent term is real, and this is true for most physical systems of practical interest, then eq.~(\ref{qft3}) reads as a weighted average of $\phi(x) \phi(x') \phi(x'') ...$ with an exponential weight factor equals to $\exp(-\mathcal{S}_E)$.

Eq.~(\ref{qft3}) looks like a correlation in statistical mechanics and the integral can be computed using standard numerical technques.

{\bf Step 3: Monte Carlo Computation}
Eq.~(\ref{qft3}) is approximated by a finite sum
\begin{equation}
\left<0\right|\phi(x) \phi(x') \phi(x'') ...\left|0\right>_E \simeq
\frac1N \sum_{k=0}^{k<N} \phi_{[k]}(x) \phi_{[k]}(x') \phi_{[k]}(x'') ...
\label{qft4}
\end{equation}

Each field configuration $\phi_{[k]}$ is generated by a random sampling from a probability distribution proportional to $\exp(-\mathcal{S}_E[\phi])$. As more terms in the sum are considered, the right hand size of eq.~(\ref{qft4}) approaches the left hand side. The numerical error in this numerical approximation can be controlled and it is usually proportional to $N^{-\frac12}$ where $N$ is the number of generated configurations.

This lattice approach to the PI can be improved to any arbtrary precision by reducing the discretization step ($a\rightarrow 0$), by considering a larger portion of space-time ($L\rightarrow \infty$), and by generating more configurations ($N\rightarrow \infty$). The extrapolation of $a$ to zero is referred to as the {\it continuum limit}.

Notice that the set of Monte Carlo configurations which appear in the sum eq.~(\ref{qft4}) does not depend on the particular correlation amplitude one is computing and therefore it can be reused for different computations as long as the system is the same (i.e. it is described by the same action). For example, in typical LQCD computations, a large numbers of configurations are generated, stored, and reused for different computations.

Because of the Wick rotation, any correlation computed using the above technique is defined in the Euclidean time, not in the Minkowsky time. Some observables are unaffected by the Wick rotation and can be reliably computed, while others require an analytical continuation back to Minkowsky time. Some phase information may be lost when combining this analytical continuation with the finite precision of the Monte Carlo method and only those observables that do not require analytical contination back to the Minkowsky time can be reliably extracted from the lattice\cite{maiani}. Luckily these include the low energy spectrum and the absolute value of matrix elements of operators between on-shell states. In LQCD typical computations include the masses of bound states such as glueballs, mesons, baryons, and matrix elements between these states.

\subsection{Algorithms}

Any Monte Carlo computation can be decomposed into three elementary steps:

{\bf Step 3.1: Markov Chain Monte Carlo (MCMC).}

\begin{figure}[t]
\begin{Verbatim}[frame=single, numbers=left, numbersep=3pt, fontfamily=helvetica, commandchars=\\\{\}, codes={\catcode`$=3\catcode`^=7}]
Algorithm: Metropolis for MCMC
Input: $\phi$, Euclidean action $\mathcal{S}$
Output: $\phi^\prime$

generate $\phi^\prime$ at random uniformly in the Hilbert space
generate a random number $z$
if $z>exp(\mathcal{S}(\phi)-\mathcal{S}(\phi^\prime,...))$ then 
    replace $\phi^\prime$ with $\phi$
return $\phi^\prime$
\end{Verbatim}
\caption{Metropolis is the simplest algorithm to generate a Markov Chain Monte Carlo (MCMC). It performs a global change of the input configuration and then an accept-reject step of that change. The latter step ensures the reversibility condition and the ergodicity of the chain. Notice that every time an update is false the output configuration is the same as the input.\label{fig:metropolis}}
\end{figure}

\begin{figure}[t]
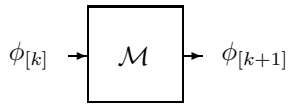

\begin{Verbatim}[frame=single, numbers=left, numbersep=3pt, fontfamily=helvetica, commandchars=\\\{\}, codes={\catcode`$=3\catcode`^=7}]
Algorithm: Gibbs sampling for MCMC
Input: $\phi$, euclidean gauge action $\mathcal{S}$
Output: $\phi^\prime$

for all lattice sites $x$ 
    store the field variable $\phi(x)$ in $\phi^{old}$
    replace the value of $\phi(x)$ with a random one (uniform in domain)
    generate a random number $z$
	if $z>exp(-\textrm{[change in action]})$ then 
            replace $\phi(x)$ with $\phi^{old}$
$\phi^\prime = \phi$
return $\phi^\prime$
\end{Verbatim}
\caption{Gibbs sampling is another algorithm to generate the MCMC. It loops over all degrees of freedom of the system and, for each of them, it performs a local change and an accept-reject of the change. If the action is local, Gibbs sampling and algorithms derived from it are more efficient than the Metropolis.
 \label{fig:gibbs}}
\end{figure}

The MCMC is a method to generate the field configurations $\phi$ by random sampling from a known probability distribution which, in this case, is proportional to $\exp(-\mathcal{S}_E[\phi])$.
\begin{center}
\setlength{\unitlength}{.02in}
\begin{picture}(75,35)(0,0)

\put(25,5){\line(0,1){25}}
\put(25,5){\line(1,0){25}}
\put(50,5){\line(0,1){25}}
\put(25,30){\line(1,0){25}}
\put(37,17){\makebox(0,0){$\mathcal{M}$}}

\put(20,17){\vector(1,0){5}}
\put(15,17){\makebox(0,0)[r]{$\phi_{[k]}$}}

\put(50,17){\vector(1,0){5}}
\put(60,17){\makebox(0,0)[l]{$\phi_{[k+1]}$}}

\end{picture}
\end{center}
The idea behind the Markov Chain is that of bulding a randomized iterative algorithm $\mathcal{M}$ so that the transition probability $P_\mathcal{M}(\phi'|\phi)$ of going from a configuration $\phi$ to a configuration $\phi'$ satisfies the following reversibility condition
\begin{equation}
e^{-S_E(\phi)}P_\mathcal{M}(\phi'|\phi) = e^{-S_E(\phi',...)}P_\mathcal{M}(\phi|\phi')
\label{mcmc}
\end{equation}

Regardless of the starting configuration $\phi_{[0]}$, the succession in $k$ is ergodic with the desired stationary distribution\cite{MCMC} $\exp(-S_E(\phi,...))$. The simplest MCMC algorithm that satisfies the reversibity condition, eq.~(\ref{mcmc}), is the Metropolis\cite{metropolis} algorithm shown in fig.~\ref{fig:metropolis}. This algorithm makes the next configuration in the chain by either picking a copy of the preceding one, or a totally new configuration generated with uniform probability in the configurations' domain. This choice is implemented as an accept-reject step that depends on the action and guarantees that the transition probability satisfies the reversibility condition.

One algorithm derived from the Metropolis is the Gibbs sampling, fig.~\ref{fig:gibbs}. It uses an accept-reject similar to the Metropolis algorithms but differs because only one field variable is updated at one time. If the action is local, and this is usually true for most physical systems, the accept-reject condition is also local, and the overall algorithm is more efficient than the Metropolis in sampling the space.

There are many other algorithms that can be used for generating the MCMC configurations and most of them are derived from either the Metropolis or the Gibbs sampling.

{\bf Step 3.2: Measure the operator.} 
This algorithm measures the desired correlation, the operator $\mathcal{O}=\phi(x)\phi(x')\phi(x'')...$, on each MCMC configuration $\phi_{[k]}$,
\begin{center}
\setlength{\unitlength}{.02in}
\begin{picture}(75,35)(0,0)

\put(25,5){\line(0,1){25}}
\put(25,5){\line(1,0){25}}
\put(50,5){\line(0,1){25}}
\put(25,30){\line(1,0){25}}
\put(37,17){\makebox(0,0){$\mathcal{O}$}}

\put(20,17){\vector(1,0){5}}
\put(15,17){\makebox(0,0)[r]{$\phi_{[k]}$}}

\put(50,17){\vector(1,0){5}}
\put(60,17){\makebox(0,0)[l]{$\phi_{[k]}(x)\phi_{[k]}(x')\phi_{[k]}(x'')...$}}

\end{picture}
\end{center}

This step is non-trivial because the operator $\mathcal{O}$ may depend on 
fields that do not appear in the PI, for example fermions propagating in a background gauge field. If this is true, the algorithm $\mathcal{O}$ requires the computation of fermion propagators. We'll provide more details in a later section.

{\bf Step 3.3: Average and Analysis.}

The final result is computed by averaging the measurements of each configuration. This average is accompanied by a statistical analysis to determine the error in the result.
\begin{center}
\setlength{\unitlength}{.02in}
\begin{picture}(75,35)(0,0)

\put(25,5){\line(0,1){25}}
\put(25,5){\line(1,0){25}}
 \put(50,5){\line(0,1){25}}
\put(25,30){\line(1,0){25}}
\put(37,17){\makebox(0,0){$\mathcal{A}$}}

\put(20,25){\vector(1,0){5}}
\put(15,25){\makebox(0,0)[r]{$\phi_{[1]}(x)\phi_{[1]}(x')\phi_{[1]}(x'')...$}}

\put(20,17){\vector(1,0){5}}
\put(15,17){\makebox(0,0)[r]{$\phi_{[2]}(x)\phi_{[2]}(x')\phi_{[2]}(x'')...$}}

\put(20,10){\vector(1,0){5}}
\put(15,10){\makebox(0,0)[r]{$\phi_{[3]}(x)\phi_{[3]}(x')\phi_{[3]}(x'')...$}}

\put(50,17){\vector(1,0){5}}
\put(60,22){\makebox(0,0)[l]{$\left<0\right|\phi(x) \phi(x') \phi(x'') ...\left|0\right>$}}
\put(60,12){\makebox(0,0)[l]{$\pm$ statistical error}}
\end{picture}
\end{center}
Estimating the stastical error is of crucial importance since it must used as a stopping condition for the MCMC. A naive estimate of the error is given by $\sigma/\sqrt{N}$ where $\sigma$ is the standard deviation of the individual measurements $\mathcal{O}_{[k]}$. However, this estimate fails when the individual measurements are not Gaussian distributed. The standard algorithm to estimate the statistical error in an average without making any assumption about the underlying distribution is the Bootstrap algorithm\cite{errors}. 

Every lattice computation is a combination of the above elementary steps as shown below. 

\begin{center}
\setlength{\unitlength}{.02in}
\begin{picture}(200,100)(0,0)

\put(65,25){\makebox(0,0)[l]{... continue MCMC ...}}
\put(50,25){\vector(1,0){10}}
\put(50,25){\line(0,1){10}}
\put(50,35){\line(1,0){30}}
\put(80,35){\line(0,1){10}}
\put(70,45){\vector(1,0){15}}
\put(92,45){\makebox(0,0){$\phi_{[3]}$}}
\put(99,45){\vector(1,0){11}}

\put(60,40){\line(0,1){10}}
\put(60,40){\line(1,0){10}}
\put(70,40){\line(0,1){10}}
\put(60,50){\line(1,0){10}}
\put(65,45){\makebox(0,0){$\mathcal{M}$}}

\put(50,45){\vector(1,0){10}}
\put(50,45){\line(0,1){10}}
\put(50,55){\line(1,0){30}}
\put(80,55){\line(0,1){10}}
\put(70,65){\vector(1,0){15}}
\put(92,65){\makebox(0,0){$\phi_{[2]}$}}
\put(99,65){\vector(1,0){11}}

\put(60,60){\line(0,1){10}}
\put(60,60){\line(1,0){10}}
\put(70,60){\line(0,1){10}}
\put(60,70){\line(1,0){10}}
\put(65,65){\makebox(0,0){$\mathcal{M}$}}

\put(50,65){\vector(1,0){10}}
\put(50,65){\line(0,1){10}}
\put(50,75){\line(1,0){30}}
\put(80,75){\line(0,1){10}}
\put(70,85){\vector(1,0){15}}
\put(92,85){\makebox(0,0){$\phi_{[1]}$}}
\put(99,85){\vector(1,0){11}}

\put(60,80){\line(0,1){10}}
\put(60,80){\line(1,0){10}}
\put(70,80){\line(0,1){10}}
\put(60,90){\line(1,0){10}}
\put(65,85){\makebox(0,0){$\mathcal{M}$}}

\put(50,85){\vector(1,0){10}}
\put(50,85){\makebox(0,0)[r]{$\phi_{[0]}$}}

\put(110,40){\line(0,1){10}}
\put(110,40){\line(1,0){10}}
\put(120,40){\line(0,1){10}}
\put(110,50){\line(1,0){10}}
\put(115,45){\makebox(0,0){$\mathcal{O}$}}
\put(120,45){\vector(1,0){20}}

\put(110,60){\line(0,1){10}}
\put(110,60){\line(1,0){10}}
\put(120,60){\line(0,1){10}}
\put(110,70){\line(1,0){10}}
\put(115,65){\makebox(0,0){$\mathcal{O}$}}
\put(120,65){\vector(1,0){20}}

\put(110,80){\line(0,1){10}}
\put(110,80){\line(1,0){10}}
\put(120,80){\line(0,1){10}}
\put(110,90){\line(1,0){10}}
\put(115,85){\makebox(0,0){$\mathcal{O}$}}
\put(120,85){\vector(1,0){20}}

\put(140,20){\line(0,1){70}}
\put(140,20){\line(1,0){10}}
\put(150,20){\line(0,1){70}}
\put(140,90){\line(1,0){10}}
\put(145,55){\makebox(0,0){$\mathcal{A}$}}

\put(150,55){\vector(1,0){15}}
\put(170,60){\makebox(0,0)[l]{$\left<0\right|\phi(x) \phi(x')...\left|0\right>$}}
\put(170,50){\makebox(0,0)[l]{$\pm$ statistical error}}

\put(30,10){\line(1,0){130}}
\put(30,10){\line(0,1){90}}
\put(30,100){\line(1,0){130}}
\put(160,10){\line(0,1){90}}

\put(25,20){\vector(1,0){5}}
\put(23,20){\makebox(0,0)[r]{irrelevant}}
\put(25,40){\vector(1,0){5}}
\put(23,40){\makebox(0,0)[r]{physical}}
\put(25,60){\vector(1,0){5}}
\put(23,60){\makebox(0,0)[r]{$L$, $N$}}
\put(25,80){\vector(1,0){5}}
\put(23,80){\makebox(0,0)[r]{$\mathcal{S}_E$, $\phi_{[0]}$}}

\end{picture}
\end{center}

Because of the way the configurations $\phi_{[k]}$ are generated, each of them retains some memory of the preceding configurations in the chain. This correlation dies out with the distance in MCMC steps between configurations. If the evaluation of the operator $\mathcal{O}$ on each individual configuration is more expensive than the elementary MCMC step $\mathcal{M}$, it may be wise to skip a number of configurations and evaluate the operator only on a subset of the total number of MCMC configurations. By skipping $c$ configurations at each step, for $c$ large enough, $\phi_{[k]}$ and $\phi_{[k+c]}$ will be sufficiently decorrelated and the procedure provides a better sampling of the integration space while saving CPU time. An empirical choice is making $c$ larger than the maximum distance between lattice sites in lattice units, $c>sL/a$. In fact, if one assumes that for a local action, at each elementary MCMC step information propagates only from one lattice site to the next, after $c$ steps, information should propagate all around the lattice, thus removing the memory of the preceding configuration.

From now on, when referring to a MCMC step, we will indicate the repetition of multiple elementary steps in order to achieve a sufficient decorrelation between effective consecutive configurations.

\subsection{Input of The Algorithms}

Any lattice computation takes the following input:

\begin{itemize}

\item
The action $\mathcal{S}_E$. The action determines the dynamics of the system. The same system may be represented by multiple actions that differ from each other because of irrelevant operators. These are high dimensional operators that can be added to the the Lagrangian and whose contribution to the action and to the correlation amplitudes vanishes in the continuum limit. Nevertheless, their contribution affects the rate of convergence of correlations to the continuum limit and can be relevant from a numerical point of view. The art of finding these operators and determining their coefficients is called {\it improvement} and is based on the work of Symanzik\cite{symanzik}. 

\item
The initial field configuration, $\phi_{[0]}$, from which the Markov Chain is started. The result of the computation should be independent from this choice because the MCMC loses memory of this initial configuration. One way to verify that this is true is by repeating the computation using different $\phi_{[0]}$.

\item
The lattice volume, represented by the parameter $L$. Ideally, one would like to perform computations close to the limit $L \rightarrow \infty$. In practice, this is not possible. A finite $L$ acts as an infrared cut-off which results in systematic errors known as finite size effects.
These effects can be quantified by repeating the computation at different values of $L$. For large $L$ boundary conditions become irrelevant but, for finite $L$, they do affect the computation. The usual choice consists of periodic boundary conditions in the hypercubic lattice topology:
\begin{equation}
\forall x, \phi(x+L)\stackrel{def}{=} \phi(x)
\end{equation}
For fermions often antiperiodic boundary conditions are adopted
\begin{equation}
\forall x, \psi(x+L)\stackrel{def}{=} -\psi(x)
\end{equation}
For typical LQCD computations $L\simeq 32a \simeq 2fm$ and often $L$ is chosen larger in the time direction than in the space direction. Empirical results indicate that for $L>5/m_\pi$ the corresponding systematic error is less than 1\%.

\item
The number of MCMC steps $N$. This number is limited by the total computing time available. For different operators, $\mathcal{O}$, the Monte Carlo integration converges at different rates with $N$. For typical LQCD computations $N\simeq 1000$.

\item
Physical parameters of the Lagrangian, such as masses of elementary particles, $m$, and coupling constants, $g$. 
Particles with masses $m<1/L$ and $m>1/a$ have propagators that, in Euclidean space, exhibit a correlation length larger then the lattice size $L$ and smaller than the lattice spacing $a$, respectively. Elementary particles with these masses cannot be put on the lattice in a naive way. The standard approach for light fermions consists of performing the computation with non physical values for the masses (in the allowed region $1/L<m<1/a$) and then extrapolating the results to the physical values $m\rightarrow \bar m < 1/L$. This extrapolation is called {\it chiral extrapolation}. For heavy fermions the extrapolation is just one of the possible approaches and different solutions are discussed in a later section.

\item
Irrelevant parameters. One has the freedom to add irrelevent operators to the Lagrangian, i.e. operators whose contribution to the action vanishes in the continuum limit. Some choices for the coefficients of these operators are better than others because they improve the rate of convergence of the correlation amplitudes to the continuum limit. The values of these parameters do not affect the result of the computation but they do affect how fast one gets to the result within the required precision.

\end{itemize}

Most notably, the value of $a$ is not a direct input of any lattice computation.
All the physical input parameters (masses and coupling constants) are bare parameters that, at constant physics, depend on the lattice spacing. 
Because of this implicit dependence, the choice of the coupling constant (which we'll refer to as $g$) is equivalent to setting the value of $a$. 
In general the relation betwen $g$ and $a$ is described by the Renormalizion Group Equation (RGE).

Because the RGE is often only known perturbatively, one does not exactly know {\it a priori} the value of $a$ in a lattice computation. Since every lattice quantity is computed in units of $a$, any error on $a$ introduces an uncertainty in those quantities with dimension different from zero. For example the energy spectrum is in units of $1/a$. This problem is solved by computing ratios of masses and/or matrix elements that cancel any explicit dependence on $a$. An equivalent approach used in LQCD consists of measuring $a$ by comparing the pion mass, $m_\pi$, computed from the lattice in lattice units with the physical pion mass and using this value to convert other quantities in physical units. We will see in the next section how this procedure is equivalent to choosing a renormalization condition.

The RGE indicates that the contiuum limit $a\rightarrow 0$ corresponds to a fixed point of the theory $g \rightarrow g^\ast$ (for example for asymptotically free\cite{afreedom} theories $g^\ast=0$) therefore the continuum limit is realized numerically by tuning $g$ closer to the fixed point $g^\ast$. Ideally the output of the computation should plateau as this limit is approached. In LQCD a typical example is the computation of the mass of the scalar over the vector meson for fixed bare quark masses.

\subsection{Regularization and The Continuum Limit}

On one hand the lattice provides a regularization scheme for the continuum PI. On the other hand the lattice formulation in the continuum limit provides a definition of the continuum PI. We wish to clarify the meaning of the concept of regularization and renormalization in the lattice paradigm.

We start by making two observations\cite{rabin,jaffe}:

\begin{itemize}
\item One can never measure the value of a continuum field $\phi(x)$
at every point in space-time. One can only measure 
its integral over the test function that corresponds to a physical detector, 
which has a finite extension. 
Therefore, there are mathematical reasons to require that a continuum field be defined in the space of distributions. 
In particular, one assumes that a particle can be localized to any arbitrary precision and that the corresponding 
field configuration can be a delta function $\delta(x)$.

\item One wants to model the short distance physics by
introducing a Lagrangian density that contains only local (contact)
interactions. Any non-trivial Lagrangian contains terms that are products or powers of fields.
\end{itemize}

These two observations are incompatible because the product or power of fields is not a well-defined quantity when a field is a delta function. 
The role of regularization is that of defining this product.

There is a physical reason behind this problem: if one only knows the field via a finite size detector, with a spatial resolution limited to $\bar a$, then any field fluctuation on a scale smaller than $\bar a$ should not be part of the model. This is why the model itself forces one to introduce some kind of cut-off $a < \bar a$. The effect of modes with length scale smaller then $a$ is encoded in the value of the bare coupling constants that one puts in the model. Any change in $a$ implies a change in the modes that contribute to the bare coupling constants and, therefore, their values have to be scaled accordingly in order to describe the same physical system.

The easiest way to regularize a distribution $\delta(x)$ is by replacing it with some localized function with finite support, for example
\begin{equation}
\delta(x) \rightarrow \delta_a(x)=\frac1a [ \theta(x+a/2)-\theta(x-a/2) ]
\end{equation}
This makes the product of distributions $\delta^n_a(x)$ well defined.

Let's consider now a 1D scalar theory defined in the interval $[0,L]$. Its most general correlation amplitude, defined in terms of the PI, is
\begin{equation}
\langle 0 | 
\phi (x) \phi (x') \phi(x'') ...
| 0 \rangle \stackrel{def}{=}
 \int [ {\textrm d} \phi ] 
F[\phi(x), g] 
\label{PI1}
\end{equation}
where $F[\phi(x),g]$ is the integrand
\begin{equation}
F[\phi(x),g] \stackrel{def}{=} \phi (x) \phi (x') \phi(x'') ...
e^{-{\mathcal{S}}_{\text{E}}[\phi,g]}
\label{PI2}
\end{equation}
and $g$ is a coupling constant that appears in the action.
The specific form of the action, $S_E[\phi,g]$ is unimportant but we assume that the action contains an interaction term of the form
\begin{equation}
g \int \phi^n(x) \text{d}x
\label{interactionterms}
\end{equation}
with $n>2$. This makes $F[\phi,g]$ a non-trivial function of $\phi$.

\begin{figure}[t]
\begin{center} 
\begin{tabular}{ccccc}
\includegraphics[width=3cm]{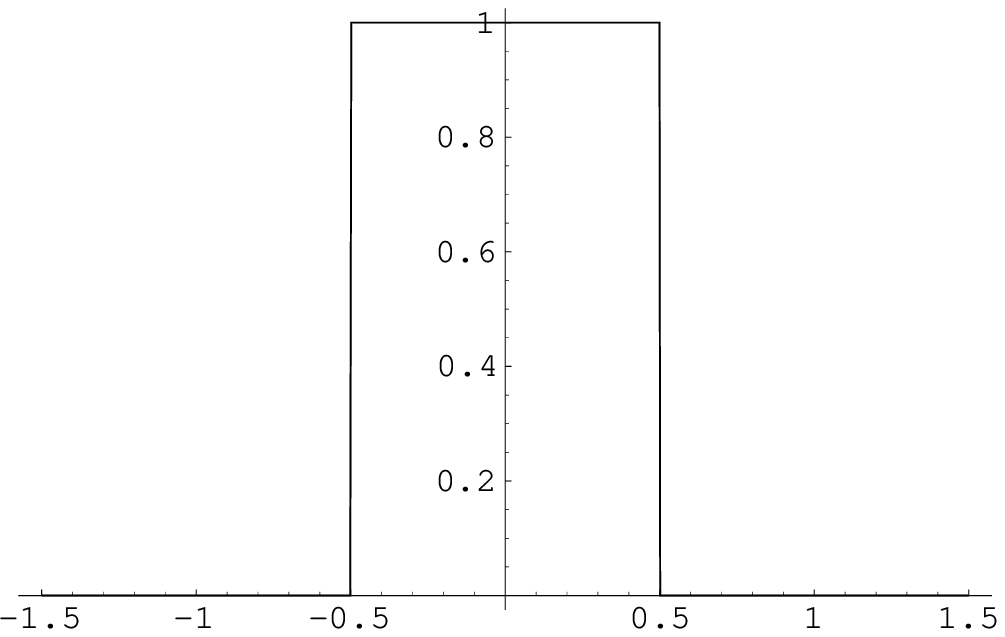} & \hfill &
\includegraphics[width=3cm]{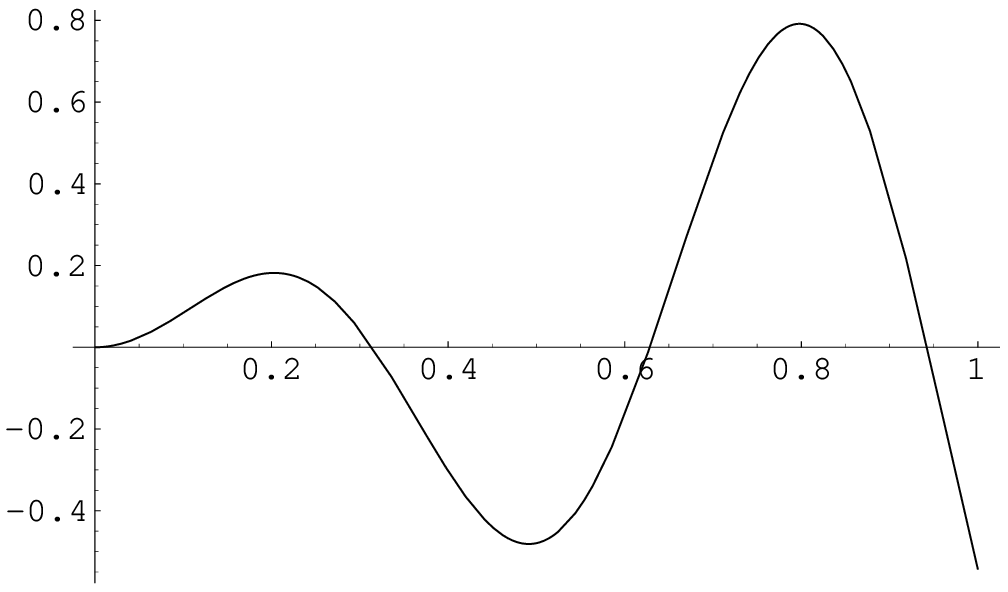} & \raisebox{1.2cm}{$\rightarrow$} &
\includegraphics[width=3cm]{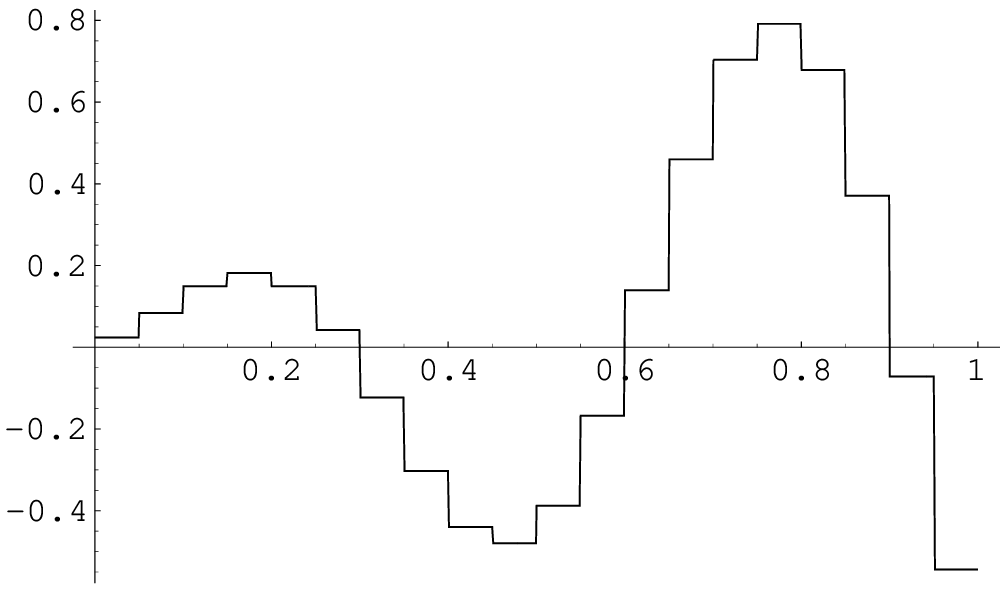} 
\end{tabular}
\end{center}
\caption{[left]:Example of possible regularization for the delta
function. The $x$ axis is in units of $a$, the $y$ axis is in units of
$1/a$. [center-right]:Example of a continuum field configuration $\phi$ and its approximation with a linear combinations of regularized delta functions. \label{delta1}}
\end{figure}

\def\imgfar{\raisebox{-5mm}{
\includegraphics[width=2.5cm]{fa00.eps} }
}
\def\imgfaq{\raisebox{-5mm}{
\includegraphics[width=2.5cm]{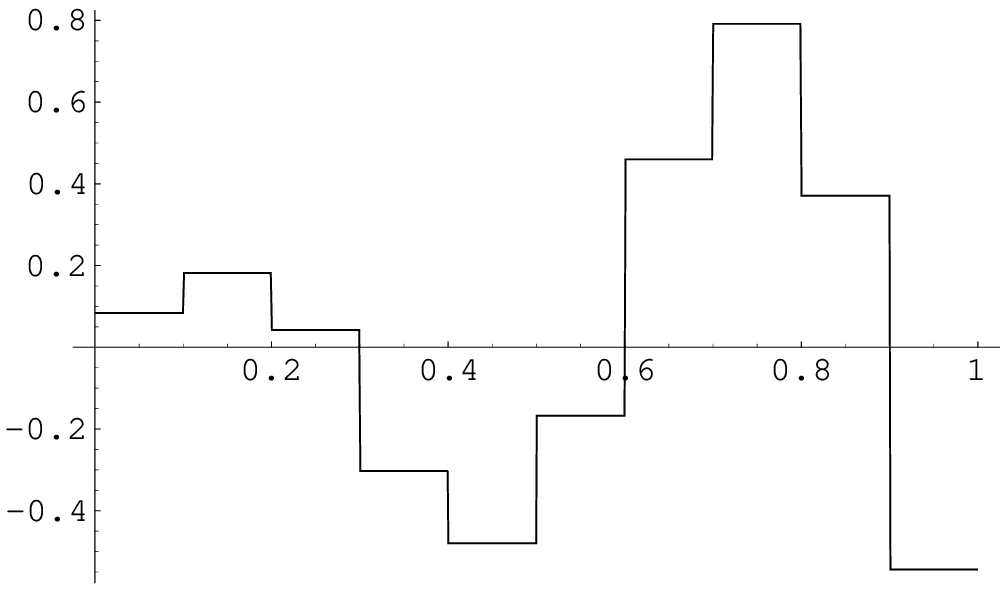} }
}
\def\imgfaw{\raisebox{-5mm}{
\includegraphics[width=2.5cm]{fa20.eps} }
}
\def\imgfae{\raisebox{-5mm}{
\includegraphics[width=2.5cm]{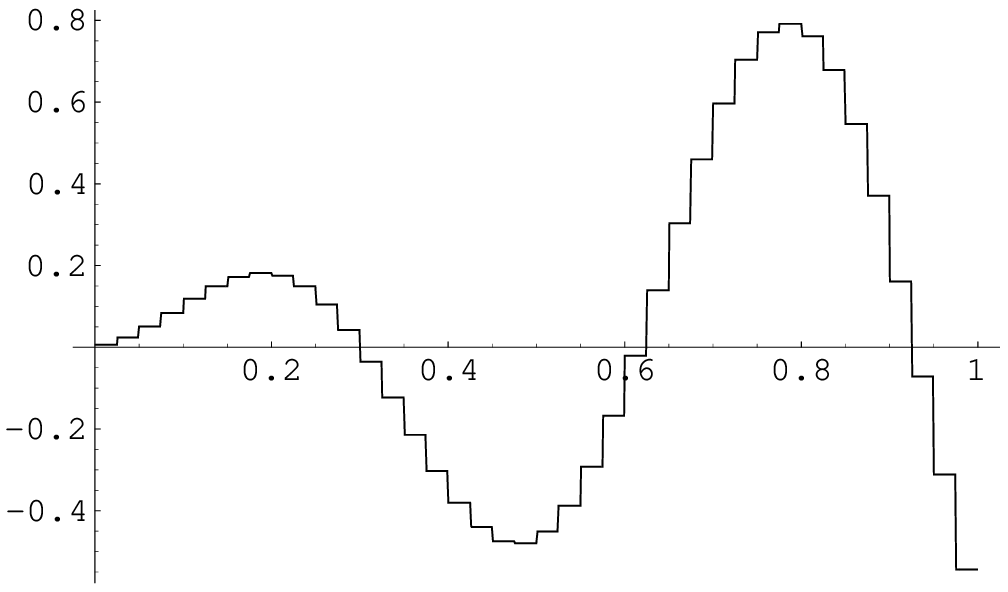} }
}
\def\imgfbr{\raisebox{-5mm}{
\includegraphics[width=2.5cm]{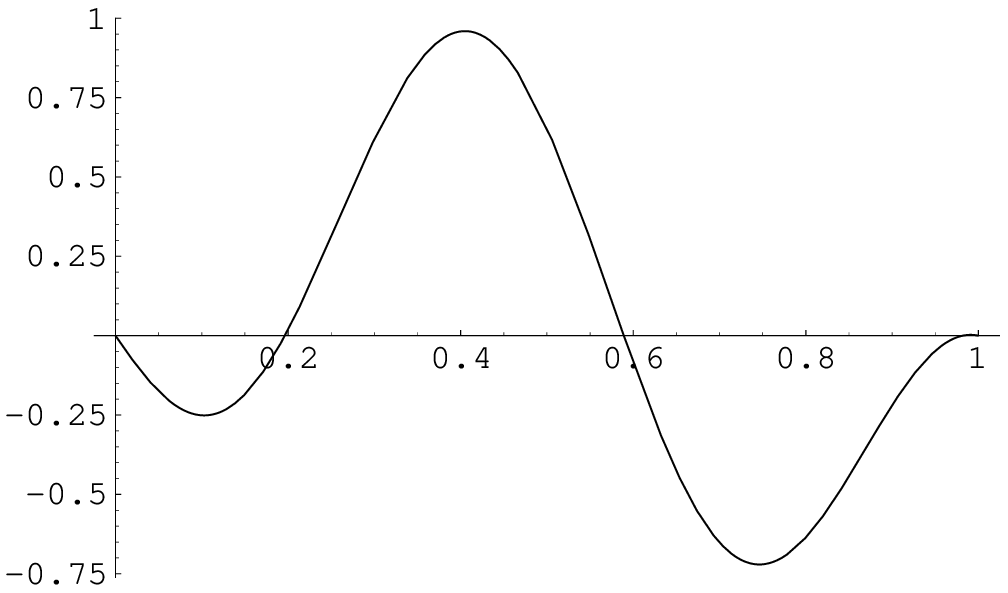} }
}
\def\imgfbq{\raisebox{-5mm}{
\includegraphics[width=2.5cm]{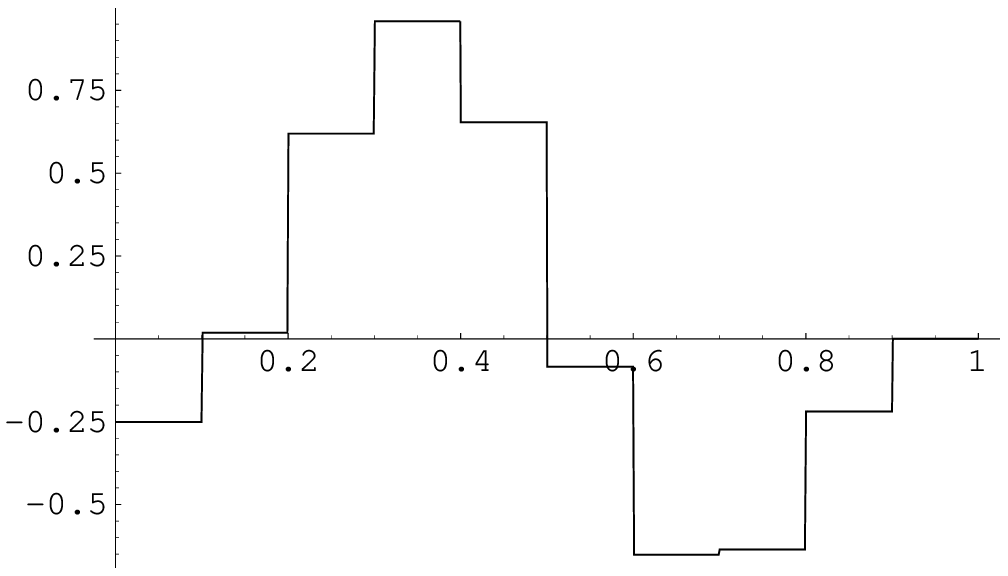} }
}
\def\imgfbw{\raisebox{-5mm}{
\includegraphics[width=2.5cm]{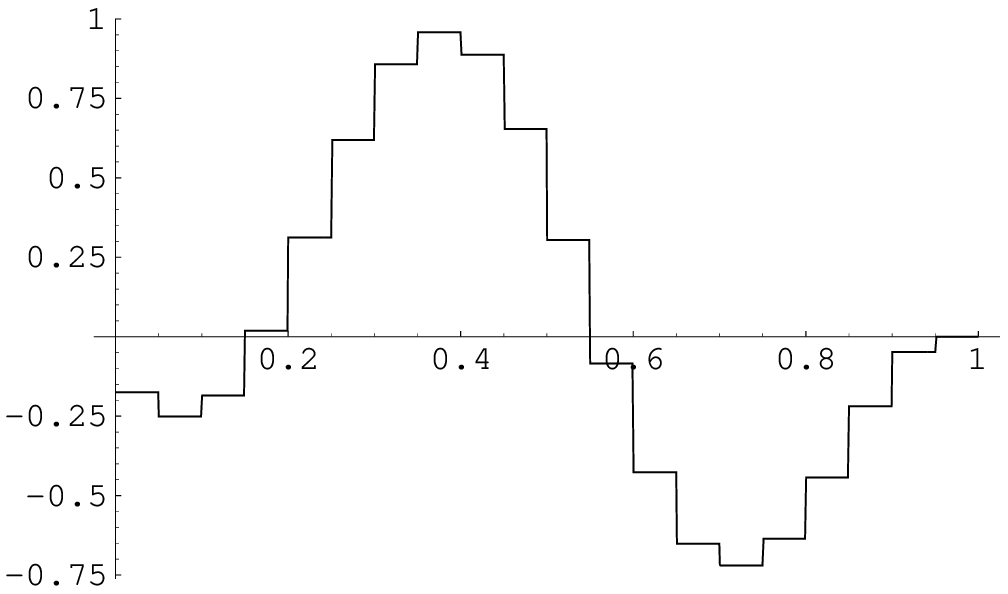} }
}
\def\imgfbe{\raisebox{-5mm}{
\includegraphics[width=2.5cm]{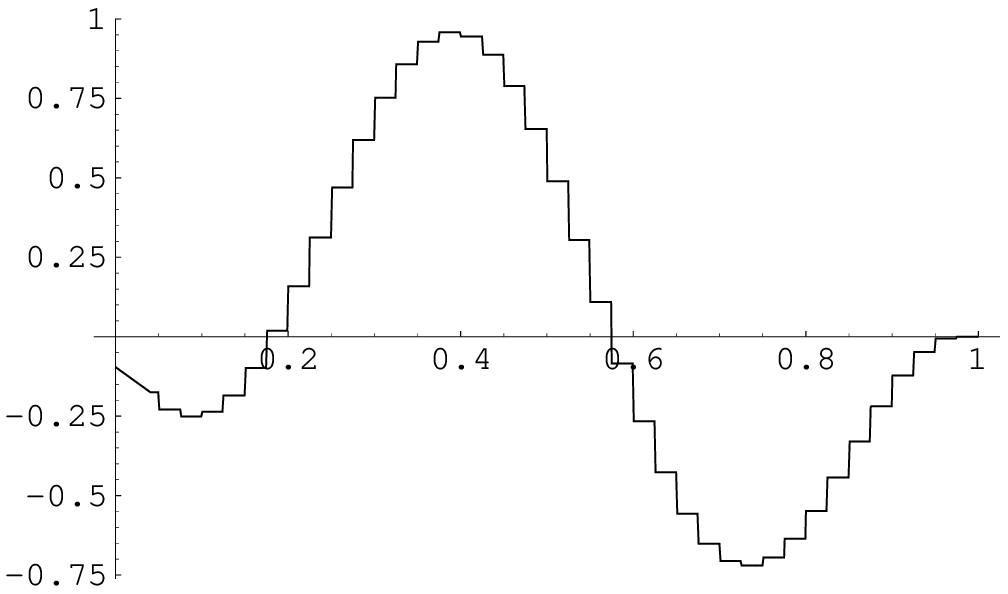} }
}
\def\imgftr{\raisebox{-5mm}{
\includegraphics[width=2.5cm]{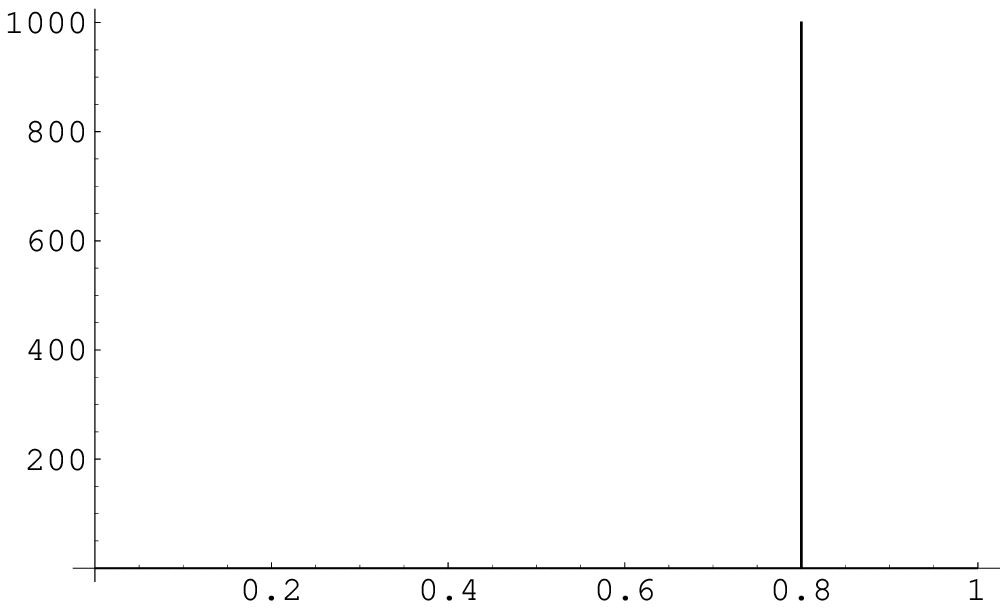} }
}
\def\imgftq{\raisebox{-5mm}{
\includegraphics[width=2.5cm]{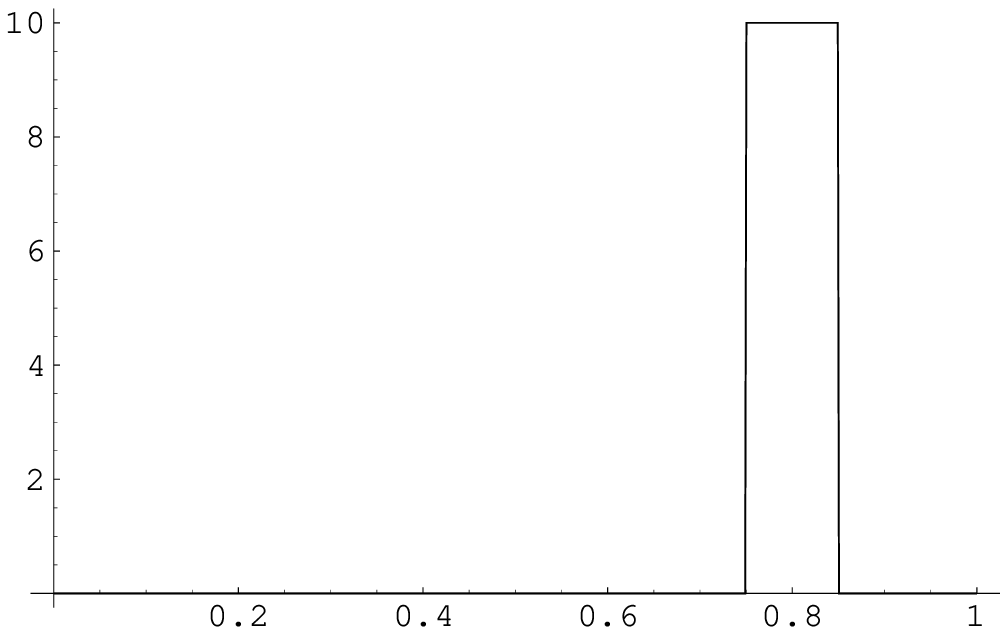} }
}
\def\imgftw{\raisebox{-5mm}{
\includegraphics[width=2.5cm]{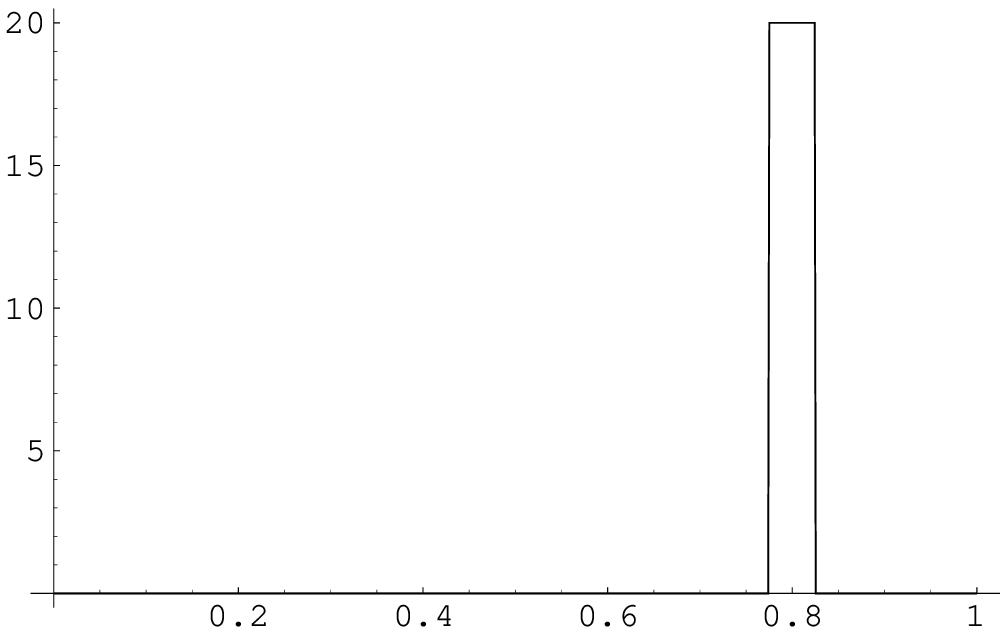} }
}
\def\imgfte{\raisebox{-5mm}{
\includegraphics[width=2.5cm]{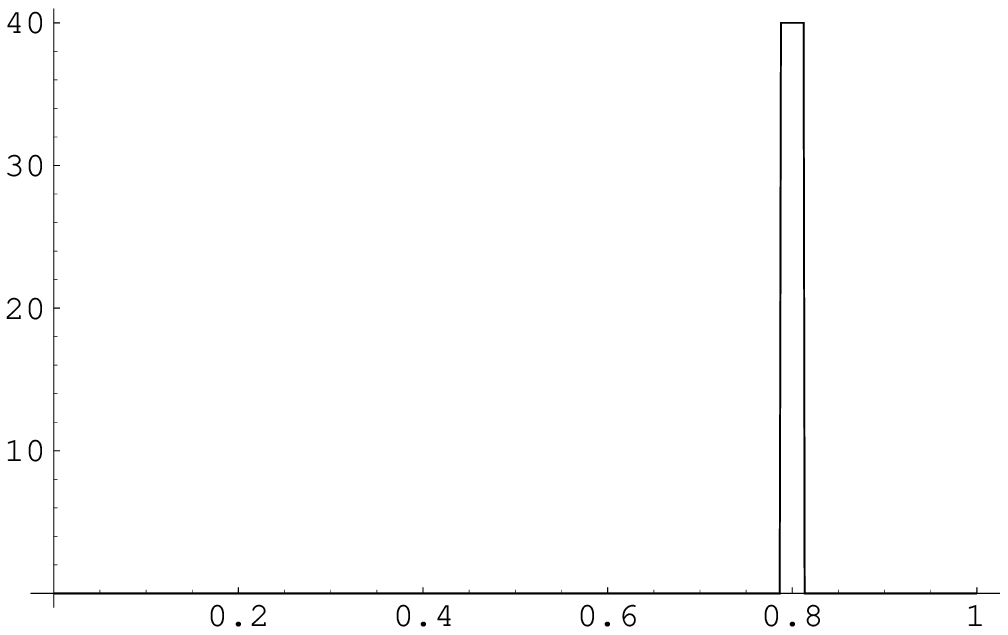} }
}

\begin{figure}
\noindent For $a=0.1$ ($K=10$):
\begin{eqnarray}
&&\int^{(a)} [ {\textrm d} \phi ] F[\phi,g] \stackrel{def}{=}
\int \text{d}\phi_0 ... \int \text{d}\phi_{9} F[\phi,g] = \label{main10} \\
&& F[\imgfaq] + F[\imgfbq] + ... + F[\imgftq] + ...\nonumber
\end{eqnarray}
For $a=0.05$ ($K=20$):
\begin{eqnarray}
&&\int^{(a)} [ {\textrm d} \phi ] F[\phi,g] \stackrel{def}{=}
\int \text{d}\phi_0 ... \int \text{d}\phi_{19} F[\phi,g] = \label{main11}\\
&& F[\imgfaw] + F[\imgfbw] + ... + F[\imgftw] + ...\nonumber
\end{eqnarray}
For $a=0.025$ ($K=40$):
\begin{eqnarray}
&&\int^{(a)} [ {\textrm d} \phi ] F[\phi,g] \stackrel{def}{=}
\int \text{d}\phi_0 ... \int \text{d}\phi_{39} F[\phi,g] = \label{main12}\\
&& F[\imgfae] + F[\imgfbe] + ... + F[\imgfte] + ...\nonumber
\end{eqnarray}
And at the continuum limit, $a\rightarrow 0$ ($K \rightarrow \infty$)
\begin{eqnarray}
&&\int [\text{d}\phi] F[\phi,g] \stackrel{def}{=}  \lim_{a\rightarrow 0} 
\underbrace{\int \text{d}\phi_0 ... \int \text{d}\phi_{K-1}}_{K \simeq L/a} F[\phi,g] = \label{main13}\\
&& F[\underbrace{\imgfar}_
{\int \phi^n(x) \text{d}x \text{ is finite}}] 
+ F[\underbrace{\imgfbr}_
{\int \phi^n(x) \text{d}x \text{ is finite}}] + ... 
+ F[\underbrace{\imgftr}_
{\int \phi^n(x) \text{d}x \rightarrow \infty}]+ ...\nonumber  
\end{eqnarray}
\caption{Example of lattice regularization of the PI\label{main1}}
\end{figure}

Lattice regularization is the way to regularize the integral~(\ref{PI1}) 
by approximating the fields with sums of regularized delta functions.
\begin{equation}
\phi (x)\simeq \phi ^{\text{latt}}(a,x)\stackrel{def}{=} \sum_{k=0}^{K-1}\phi_k\delta_a (x-ka)  \label{expansion0}
\end{equation}
where 
\begin{equation}
\phi _k\stackrel{def}{=} \frac1a \int_{ka}^{ka+a} \phi(x) \text{d}x 
\end{equation}
This is equivalent to discretizing the space-time on which the fields are defined (as shown in fig.~\ref{delta1}).

After discretization, for every finite $K=L/a$,
\begin{equation}
\int^{(a)} [ {\textrm d} \phi ] F[\phi(x),g] \stackrel{def}{=}
\underbrace{\int \text{d}\phi_0
\int \text{d}\phi_1 ...
\int \text{d}\phi_{K-1}}_{K\simeq L/a} F[\phi ^{\text{latt}}(a,x),g] + O(a)
\label{reg1}
\end{equation}

Here we used the upper index $(a)$, which identifies the regularized PI  with lattice spacing set to $a$.

Divergences associated with the limit $a\rightarrow 0, K\rightarrow \infty$ (at $L=Ka=$constant) are called ultraviolet, while those associated with the limit $K,L\rightarrow \infty $ (at $a=L/K=$constant) are called infrared.

Fig.~\ref{main1} shows, in a schematic way, how the path integral, eq.~(\ref{main13}), can be approximated by finite multidimensional integrals, eqs.~(\ref{main10}-\ref{main12}), and the fields are defined on a lattice. For each finite $a$,
the ``sum over the paths'' (eqs.~(\ref{main10}-\ref{main12})) is well defined since all divergences that may appear can be absorbed in the normalization of the integration measure and, for each path $\phi$, the functional $F[\phi,g]$ is finite. 
In the limit $a\rightarrow 0$ those configurations that correspond to a localized particle become more and more peaked and 
approach a Dirac $\delta(x)$ function (eq.~(\ref{main13})). Since the integrand $F[\phi(x),g]$ is non-linear, the integrand diverges on those configurations $\phi(x) \simeq \delta(x)$.

In order to have a well defined limit $a \rightarrow 0, K \rightarrow \infty$ (at $L=Ka=$constant) one must require that the result of the regularized path integral is independent from $a$. The only way to do it is to making the field normalization and the coupling constant (the $g$ of eq.~(\ref{interactionterms})) dependent on $a$
\begin{equation} 
g \rightarrow g_R(a,\Lambda) 
\label{ga}
\end{equation} 
(the constant $\Lambda$ must be introduced because, in general, $a$ and $g$ 
do not have the same dimensions). 
This makes the physics independent by the lattice scale $a$.

One does this by choosing a particular correlation amplitude (identified by the functional integrand $F[\phi,g]$) and imposing the contraint
\begin{equation}
\frac{\text{d}}{\text{d}a} \left[ \underbrace{
\int \text{d}\phi_0 \int \text{d}\phi_1 ...
\int \text{d}\phi_{K-1}}_{K \simeq L/a} F[\phi(x), g_R(a,\Lambda)] \right] \simeq 0
\label{RGE3}
\end{equation}
Eq.~(\ref{RGE3}) is the RGE for a lattice regularized theory and it determines the behavior (the running) of $g_R(a)$. The appearance of $\Lambda$ is called {\it dimensional transmutation}.

$\Lambda$ is the typical length scale of the physics being studied. 
This scale is in nature and there is no freedom to change it. 
For QCD, for example, it is best determination is from the LQCD scaling of the coupling constant~\cite{lambdaqcd}, $\Lambda_{\text{QCD}}^{\bar{MS}} \simeq 259(1)(20)\textrm{MeV}$ ($1/\Lambda_{\text{QCD}}\simeq 1$fm).

Notice that this procedure of defining the limit $a \rightarrow 0$ cannot be carried out for an arbitrary theory since there may be more sources of divergence than coupling constants. If this limit can be defined the theory is said to be renormalizable, if not, $a$ must be kept finite and the theory should be considered as an effective theory.

We distinguish between the {\it bare} parameters that
appear in the regularized Lagrangian (for a finite value of the
cut-off, $a$) and the {\it dressed} or {\it renormalized}
parameters that are measured by actual experiments. If one
takes the limit $a \rightarrow 0$, the bare parameters lose
any physical meaning and one must carefully define the renormalized
ones (one is said to choose a prescription). 
If one is happy with keeping the cut-off small but finite one is
allowed to identify the renormalized and the bare parameters, because
these can now be measured. This approach is known as the 
Kadanoff-Wilson approach to renormalization.

In LQCD eq.~(\ref{RGE3}) is realized numerically. One repeats the computation of the same quantity, for example the pion mass, $m_\pi$, for different values of the coupling constant $g'$, $g''$, $g'''$, etc. thus obtaining $a'\bar m_\pi$, $a''\bar m_\pi$, $a'''\bar m_\pi$, etc. where $\bar m_\pi$ is the physical pion mass. By comparing the lattice results in lattice units with the physical pion mass one can obtain the value of $a$ that corresponds to the input values of $g$. From the plot one determines $g(a)$ and identifies the fix point $g^\ast$. The same procedure can be carried on any physical quantity and it should be lead to the same running of $g$, up to corrections in $a$.

Since we will never be able to probe the physical world at every length scale,
every quantum field theory should be considered an effective theory.

\subsection{Lattice Regularization vs Momentum Cut-off}  

A continuum field $\phi$ can be expanded into Laplace components as
\begin{equation}
\phi (x)=\sum_{n=0}^\infty b_n e^{ip_nx}  \label{expansion1}
\end{equation}
where 
\begin{equation}
p_n\stackrel{def}{=} \frac{2 \pi n}L; \qquad
b_n\stackrel{def}{=} \frac 1{2\pi }\int_{0}^L\phi (x)e^{-ip_nx}dx
\end{equation}

\begin{figure}[t]
\begin{center} 
\begin{tabular}{ccc}
\includegraphics[width=4cm]{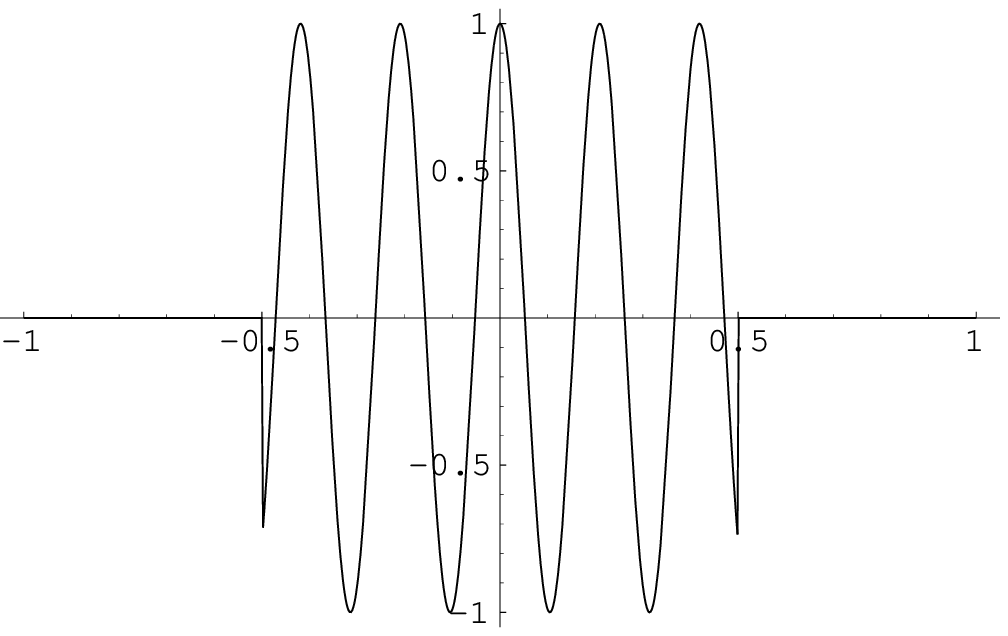} &
\includegraphics[width=4cm]{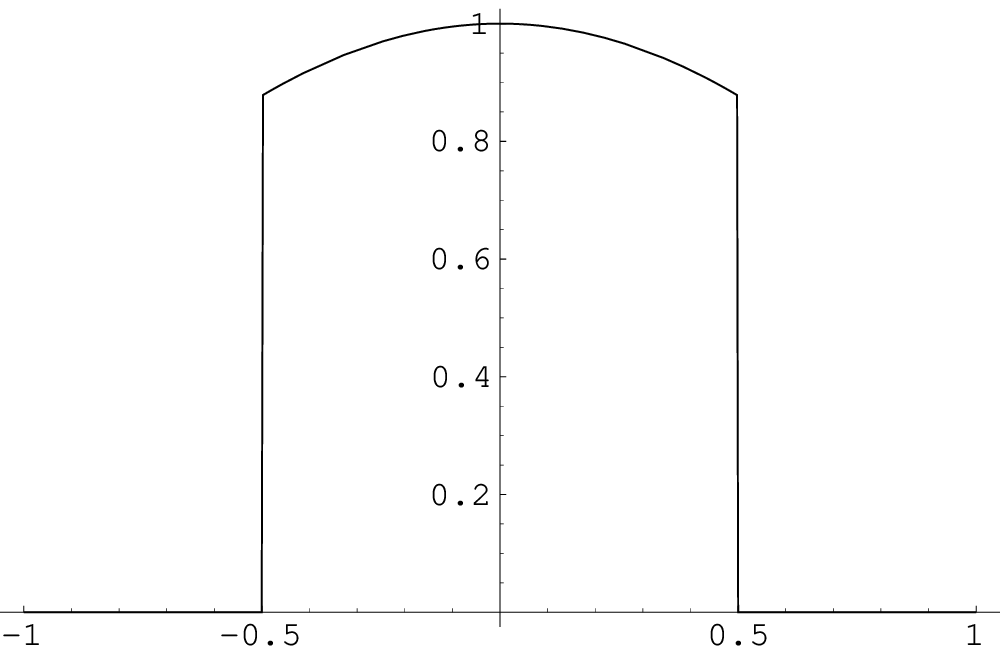}
\end{tabular}
\end{center}
\caption{Different behavior of the integrand of $b_n'$ for high
frequency modes (left) and low frequency modes (right) respectively. 
The $x$ axis is in units of $a$. \label{waves}}
\end{figure}

Similarly a lattice field, eq.(\ref{expansion0}) can also be expandend in Laplace components and we obtain
\begin{equation}
\phi ^{\text{\text{latt}}}(a,x)=\sum_{n=0}^\infty b_n^{\prime }e^{ip_nx}
\end{equation}
where 
\begin{equation}
b_n^{\prime }=\frac 1{2\pi }\sum_{k=0}^{K-1}\left[ \phi
_k\int_{0}^L\delta_a (ka-x)e^{-ip_nx}dx\right] 
\end{equation}
It becomes evident that for $p_n>1/a$ the integrand oscillates quickly and
the corresponding integral, $b_n^{\prime }$, is small; while for $%
p_n<1/a$ the integral is almost constant and approximately equal to $%
e^{-ip_nka}$, therefore $b_n^{\prime }\simeq b_n$. 
The different behaviors of the integrand are shown in figure \ref{waves}.
This proves that eq.(\ref{expansion1}) can be written as 
\begin{equation}
\phi (x)\simeq \phi ^{\text{\text{latt}}}(a,x)\simeq \phi ^{\text{co}}(a,x)\stackrel{def}{=}
\sum_{n=0}^\infty \theta \left(\frac{1}{a}-p_n\right) b_ne^{ip_nx}
\end{equation}

The superscripts ``latt'' and ``co'' are used to identify the lattice and 
and the cut-off regularization schemes, respectively.

In other words the lattice cut-off $a$ is equivalent to a momentum cut-off $p_{\text{max}}<1/\bar a$, an ultraviolet cut-off.
Therefore the lattice and the momentum cut-off are alternative but equivalent ways to regularize the PI.

Note that $b_0$ is the mean value of $\phi(x)$ and  $p_1=2\pi/L$ represents the minimum energy/momentum mode that can
propagate on a finite unidimensional volume of length $L$. This is a lower limit, an infrared cut-off.

\index{regularization!lattice}
\index{regularization!momentum cut-off}
\index{regularization!Pauly-Villars}
\index{regularization!dimensional}

\section{Pure Gauge Theories}

\subsection{Action}

We consider a pure gauge theory and we restrict the gauge group $\mathcal{G}$ to $U(1)$ and/or $SU(n)$. 
In order to be able to probe the gauge field we assume to have a single particle $\psi$ of infinite mass and unit charge that we can move on the lattice as we please.

The Aharonov-Bohm experiment\cite{aharonov} teaches us that the gauge field is not directly measurable but if we move the test particle $\psi$ from one point $x$ to a point $x'$ along a path $C$, the test charge acquires a phase that can be measured via interference experiments
\begin{equation}
\psi(x')=\exp\left(ig \int_C A_\mu(x) \textrm{d}x^\mu \right) \psi(x)
\label{ahronov}
\end{equation}
where $A^\mu(x)$ is the gauge field in the continuum space. On a lattice the shortest possible path is the link connecting two neighbor sites $x$ and $x+\hat\mu$ ($+\hat\mu$ here indicates a positive vector in direction $\mu$ and length equal to the lattice spacing $a$) therefore the elementary lattice gauge degrees of freedom are not $A_\mu$ but 
\begin{equation}
U(x,\mu) \stackrel{def}{=} \exp\left(ig\int_{x}^{x+\hat\mu} A_\mu(x) \textrm{d}x^\mu \right) \simeq 1 + iag A_\mu(x+\frac12\hat\mu) + O(a^2)
\label{defu}
\end{equation}
For later convenience we also define
\begin{equation}
U(x,-\mu) \stackrel{def}{=} \exp\left(ig\int_{x+\hat\mu}^{x} A_\mu(x) \textrm{d}x^\mu \right) \simeq 1 - iag A_\mu(x-\frac12\hat\mu) + O(a^2)
\label{defu2}
\end{equation}
From the definition it follows that $U(x,-\mu)=U^\dagger(x-\hat\mu,\mu)$.

For the rest of this section the gauge links $U(x,\mu)$ will replace our generic template field $\phi(x)$. The index $\mu$ is an integer that labels the vector component of the gauge field.

Under a local gauge transformation $\psi(x)\rightarrow V(x) \psi(x)$ the field $U$ transforms as
\begin{equation}
U(x,\mu) \rightarrow V(x)U(x,\mu)V^{-1}(x+\hat\mu)
\end{equation}
Any path $C$ on the lattice can be identified with a set of links $U(x_{[i]},\mu_{[i]})$ such that 
\begin{equation}
\exp{\int_C A^\mu(x) dx_\mu} \simeq \prod_i U(x_{[i]},\mu_{[i]})
\end{equation}

The most local gauge invariant one can build is called the plaquette and it is a product of 4 links in a loop
\begin{equation}
P(x,\mu,\nu) \stackrel{def}{=} Tr \left[ U(x,\mu) U(x+\hat\mu,\nu) U(x+\hat\mu+\hat\nu,-\mu) U(x+\hat\nu,-\nu) \right]
\label{plaquette}
\end{equation}
Here are some examples of plaquettes:
\begin{center}
\setlength{\unitlength}{.02in}
\begin{picture}(75,75)(0,0)
\multiput(25,25)(10,0){4}{\circle*{3}}
\multiput(25,35)(10,0){4}{\circle*{3}}
\multiput(25,45)(10,0){4}{\circle*{3}}
\multiput(25,55)(10,0){4}{\circle*{3}}

\multiput(28,27)(10,0){4}{\circle*{2}}
\multiput(28,37)(10,0){4}{\circle*{2}}
\multiput(28,47)(10,0){4}{\circle*{2}}
\multiput(28,57)(10,0){4}{\circle*{2}}

\multiput(31,29)(10,0){4}{\circle*{2}}
\multiput(31,39)(10,0){4}{\circle*{2}}
\multiput(31,49)(10,0){4}{\circle*{2}}
\multiput(31,59)(10,0){4}{\circle*{2}}

\put(25,25){\vector(0,1){10}}
\put(25,35){\vector(1,0){10}}
\put(35,35){\vector(0,-1){10}}
\put(35,25){\vector(-1,0){10}}

\put(35,45){\vector(0,1){10}}
\multiput(35,55)(0.3,0.2){10}{\circle*{1}}
\put(38,57){\vector(0,-1){10}}
\multiput(35,45)(0.3,0.2){10}{\circle*{1}}

\put(45,45){\vector(1,0){10}}
\multiput(55,45)(0.3,0.2){10}{\circle*{1}}
\put(58,47){\vector(-1,0){10}}
\multiput(45,45)(0.3,0.2){10}{\circle*{1}}

\put(15,30){\line(1,0){8}}
\put(14,30){\makebox(0,0)[r]{$P(x,1,2)$}}

\put(15,52){\line(1,0){18}}
\put(14,52){\makebox(0,0)[r]{$P(x,2,3)$}}

\put(70,45){\line(-1,0){10}}
\put(72,45){\makebox(0,0)[l]{$P(x,1,3)$}}

\end{picture}
\end{center}

The simplest lattice action $\mathcal{S}_E$ one can engineer using the gauge field must therefore be linear in $P(x,\mu,\nu)$, real, and invariant under the lattice rotational symmetry. Following the common notation this simplest action can be written as
\begin{equation}
\mathcal{S}_E^{gauge} \stackrel{def}{=} \frac{-\beta}{2n} Re \sum_{x,\mu\neq\nu} P(x,\mu,\nu)
\label{gaugeaction}
\end{equation}
where $n$ is the size of the gauge group ($n=1$ for $U(1)$).

Substituting eqs.~(\ref{defu}-\ref{plaquette}) in eq.~(\ref{gaugeaction}) one obtains
\begin{eqnarray}
\mathcal{S}_E^{gauge} &\simeq \frac{-\beta}{2n} Re \sum_{x,\mu\nu}&
[1+iagA_\mu(x+\frac12\hat\mu)] [1+iagA_\mu(x+\frac12\hat\nu)] \times \nonumber \\
&&[1-iagA_\mu(x-\frac12\hat\mu)] [1-iagA_\nu(x-\frac12\hat\nu)] \nonumber \\
&\simeq \frac{-\beta}{2n} Re \sum_{x,\mu\nu}&
[1+iagA_\mu - \frac{ia^2g}2 \partial_\nu A_\mu]
[1+iagA_\mu + \frac{ia^2g}2 \partial_\mu A_\nu] \times \nonumber \\
&&[1-iagA_\mu - \frac{ia^2g}2 \partial_\nu A_\mu]
[1-iagA_\nu + \frac{ia^2g}2 \partial_\mu A_\nu] \nonumber \\
&\simeq \frac{-\beta}{2n} Re \sum_{x,\mu\nu}&
1 - \frac{a^4 g^2}4 
(\partial_\mu A_\nu-\partial_\nu A_\mu+g[A_\mu,A_\nu])^2 \nonumber \\
&\simeq c + \frac{\beta}{2n} \sum_{x,\mu\nu}& 
\frac{a^4g^2}4 F_{\mu\nu}F^{\mu\nu} \label{gauge_taylor2} 
\end{eqnarray}
where $c$ is an overall irrelevant constant and eq.~(\ref{gauge_taylor2}) was derived from eq.~(\ref{gaugeaction}) via a Taylor expansion around $x+\frac12\mu+\frac12\nu$.

Notice that $\frac14 F_{\mu\nu}(x) F^{\mu\nu}(x)$ is the ordinary kinetic term for a continuum gauge field and $a^4 \sum_{x}$ is $\int \textrm{d}^4x$ in the contiuum limit. Following this analogy 
\begin{equation}
\beta = \frac{2n}{g^2}
\end{equation}
hence $g$ is interpreted as the  bare gauge coupling constant at the lattice scale $a$. For non-abelian gauge theories asymptotic freedom dictates that $g$ goes to zero when $a$ goes to zero.

Eq.~(\ref{gaugeaction}) is called Wilson gauge action\cite{Wilson:1974}
The Wilson gauge action can also be derived by direct discretization of the continuum gauge action by ignoring everything but the lowest order terms in $a$.

For large $\beta$ the trace of the average plaquette is close to $n,$ perturbative effects dominate and this results in small quantum fluctuations, and small/slow changes in the configurations generated by the MCMC algorithm. For small $\beta$ the average plaquette is close to 0, non-pertubative effects dominate which results in large/non-local quantum fluctuations, and relatively big changes in the configurations generated by the MCMC.

\subsection{Algorithms}

\begin{figure}[t]
\begin{Verbatim}[frame=single, numbers=left, numbersep=3pt, fontfamily=helvetica, commandchars=\\\{\}, codes={\catcode`$=3\catcode`^=7}]
Algorithm: Generate a random $SU(n)$ matrix
Input: $n$
Output: $A$

if $n==1$ return $exp(2\pi i uniform())$
$A=n\times n$ identity matrix
for $i=0$ to $n-2$
    for $j=i+1$ to $n-1$
        $\alpha=\pi \textrm{uniform()}$
        $\phi=2 \textrm{uniform()}$
        $\cos(\theta)=2 \textrm{uniform()}-1$
        $\sin(\theta)=\sqrt{1-\cos^2(\theta)}$
	$u^0=\cos(\alpha)$
	$u^1=\sin(\alpha)\sin(\theta)\cos(\phi)$
	$u^2=\sin(\alpha)\sin(\theta)\sin(\phi)$
	$u^3=\sin(\alpha)\cos(\theta)$
	$G=u^0+u^1\sigma^1+u^2\sigma^2+u^3\sigma^3$
	$A^\prime = A$
        for $k=0$ to $n-1$
            $A^{\prime ik}=G^{00}A^{ik}+G^{01}A^{jk}$
            $A^{\prime jk}=G^{10}A^{ik}+G^{11}A^{jk}$
	$A=A^\prime$      
return $A$.
\end{Verbatim}
\caption{General algorithm for generating a random element in a $U(1)$ and/or $SU(n)$ gauge group with uniform distribution within the group. For $SU(2)$ it uses the invertible map in $SO(3)$ and for $SU(n>2)$ it generates the matrix as product of $SU(2)$ subgroups. The function $uniform()$ is assumed to return a uniform random number in $(0,1)$.\label{Cabibbo}}
\end{figure}

Because of the locality of the Wilson gauge action, for every link $U(x,\mu)$ one can rewrite the action as 
\begin{equation}
\mathcal{S}_E^{gauge}=\frac{-\beta}{2n} Re Tr U(x,\mu) R(x,\mu) + ...
\label{gaugeaction_heatbath}
\end{equation}
where the dots represent the sum over plaquettes that do not include $U(x,\mu)$ and
\begin{eqnarray}
R(x,\mu)&=&\sum_{\nu\neq\mu} U(x+\hat\mu,\nu)U(x+\hat\mu+\hat\nu,-\mu)U(x+\hat\nu,-\nu)+ \nonumber \\
&&\phantom{\sum_{\nu\neq\mu}} U(x+\hat\mu,-\nu)U(x+\hat\mu-\hat\nu,-\mu)U(x-\hat\nu,\nu)
\end{eqnarray}
are referred to as {\it staples}. A 3D projection is represented in the image below
\begin{center}
\setlength{\unitlength}{.02in}
\begin{picture}(75,75)(0,0)
\multiput(25,25)(10,0){4}{\circle*{3}}
\multiput(25,35)(10,0){4}{\circle*{3}}
\multiput(25,45)(10,0){4}{\circle*{3}}
\multiput(25,55)(10,0){4}{\circle*{3}}

\multiput(28,27)(10,0){4}{\circle*{2}}
\multiput(28,37)(10,0){4}{\circle*{2}}
\multiput(28,47)(10,0){4}{\circle*{2}}
\multiput(28,57)(10,0){4}{\circle*{2}}

\multiput(31,29)(10,0){4}{\circle*{2}}
\multiput(31,39)(10,0){4}{\circle*{2}}
\multiput(31,49)(10,0){4}{\circle*{2}}
\multiput(31,59)(10,0){4}{\circle*{2}}

\multiput(38,37)(1,0){10}{\circle{1}}

\put(48,47){\vector(-1,0){10}}
\put(38,37){\line(0,1){10}}
\put(48,37){\line(0,1){10}}

\put(48,27){\vector(-1,0){10}}
\put(38,27){\line(0,1){10}}
\put(48,27){\line(0,1){10}}

\put(45,35){\vector(-1,0){10}}
\put(51,39){\vector(-1,0){10}}
\multiput(35,35)(0.3,0.2){20}{\circle*{1}}
\multiput(45,35)(0.3,0.2){20}{\circle*{1}}

\end{picture}
\end{center}

This makes the Gibbs sampling algorithm more efficient than the Metropolis because it is possible to change a single link $U(x,\mu)=U^{old}\rightarrow U(x,\mu)=U^{new}$ at one time without having to recompute the entire action. In fact, the accept-reject step only depends on the variation of the action given by the first term in eq.~(\ref{gaugeaction_heatbath})
\begin{equation}
\delta\mathcal{S}_E^{gauge}=\frac{-\beta}{2n}ReTr[(U^{new}-U^{old}))R(x,\mu)]
\end{equation}

There are many algorithms that are similar to the Gibbs sampling but more efficient. One of the most common is the heatbath\cite{heatbath}.

In order to make a MCMC step, whether global in the Metropolis or local in the Gibbs sampling and derived algorithms, one must be able to generate a new link at random with uniform probability in the gauge group $\mathcal{G}$.

For $\mathcal{G}=U(1)$ it is sufficient to generate a uniform random number $u\in[0,1]$ and update 
\begin{equation}
U(x,\mu)\rightarrow U'(x,\mu)=e^{2\pi i u}
\end{equation}

For $\mathcal{G}=SU(2)$ one can use the map between $SU(2)$ and $SO(3)$ (the symmetry group of a 3-sphere), generate a uniform point on a 3-sphere $(u^0, u^1,u^2,u^3)$ and map it back into $SU(2)$ via
\begin{equation}
U(x,\mu)\rightarrow U'(x,\mu)=u^0 + u^1 \sigma^1 +u^2 \sigma^2 +u^3 \sigma^3
\end{equation}
where $\sigma_i$ are the Pauli matrices.

For $\mathcal{G}=SU(n)$ and $n>2$ there is no exact technique but a common recursive technique\cite{cabibbo} consists of
\begin{equation}
U(x,\mu)\rightarrow U'(x,\mu)=\prod_{i<j} G_{ij} 
\end{equation}
where $G_{ij}$ are random $SU(2)$ matrices that acts on the $ij$ subgroup of $SU(n)$.

The general algorithm is reported in fig.~\ref{Cabibbo}.

\subsection{Example: Quark-Antiquark Potential}

\begin{figure}[t]
\begin{Verbatim}[frame=single, numbers=left, numbersep=3pt, fontfamily=helvetica, commandchars=\\\{\}, codes={\catcode`$=3\catcode`^=7}]
Algorithm: Compute the static quark-antiquark potential
Input: $\beta$, size of gauge group $n$, number of MCMC steps $N$
Output: $V(r)$

Create local array $V(r)$ and initialize it to zero

for each lattice site $x$
    for each direction $\mu$
        set $U(x,\mu)$ to a random element of the gauge group $SU(n)$

for $c=1$ to $N$
    replace $U$ with the next configuration in the MCMC (use $\beta$)
    for each lattice site $x$
        for each rectangular path $r\times t$ starting in $x$
            compute $P^{r\times t}(x)$, the product of links along the path
	    compute $v=-log(P^{r\times t}(x))/(t\times L^4)$
            add $v$ to $V(r)$
return $V(r)$
\end{Verbatim}
\caption{Example of algorithm to compute the static quark-antiquark potential. Notice the role of steps 7-9 is to create the initial configuration, step 11 loops over the Markov chain, step 12 creates the next configuration in the chain (using Metropolis, Gibbs sampling or other algorithm), steps 13-16 measure the operator, and step 17 averages the results.\label{fig:algotrithmvr}}
\end{figure}

According to Quantum Mechanics, a state consisting of one static quark and one static antiquark separated by a distance $r$ evolves in time by aquiring a phase $e^{iHt}$ where $H$, the Hamiltonian, is equal to the static potential between the quarks $H=V(r)$. In fact, kinetic contribution to $H$ is zero because the quarks are static.

\begin{figure}[t]
\begin{center}
\includegraphics[width=4in]{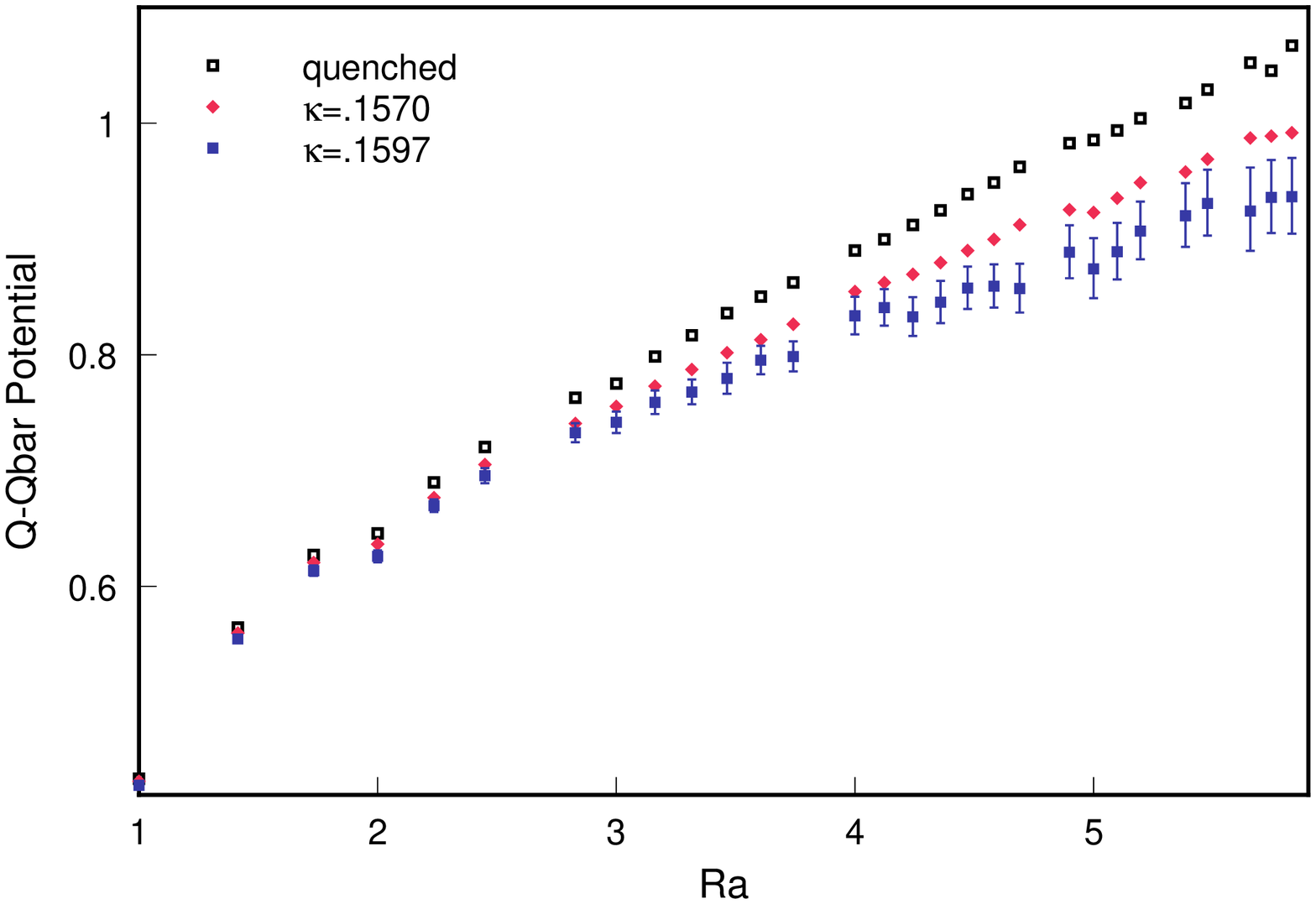}
\end{center}
\caption{The static quark potential for quenched and two flavor dynamical QCD for different masses of the sea quarks$^{22}$. Results are in units of the lattice spacing.\label{staticpotential}}
\end{figure}

On a Euclidean lattice the same state evolves according to $e^{-V(r)t}$, therefore the potential $V(r)$ can be determined by computing the lattice expectation value of the operator that corresponds to this system.

The quark and the anti-quark have to be created at the same point, separated at distance $r$, evolve for a time $t$ and then reunite and annihilate. Due to the fact that an antiquark is nothing other that a quark propagating backwards in time, the operator that corresponds to this system is the product of links around a square path of size $r\times t$
\begin{eqnarray}
P^{r\times t}(x)&\stackrel{def}{=}& 
U(x,\mu)U(x+\hat\mu,\mu)...U(x+(r-1)\hat\mu,\mu) 
\times \nonumber \\
&&U(x+r\hat\mu,\nu)U(x+r\hat\mu+\hat\nu,\nu)...U(x+r\hat\mu+(t-1)\hat\nu,\nu) 
\times \nonumber \\
&&U(x+r\hat\mu+t\hat\nu,-\mu)U(x+(r-1)\hat\mu+t\hat\nu,-\mu)...U(x+\hat\mu+(t-1)\hat\nu,-\mu) 
\times \nonumber \\
&&U(x+t\hat\nu,-\nu)U(x+(t-1)\hat\nu,-\nu)...U(x+\hat\nu,-\nu) 
\end{eqnarray}

The static potential can therefore be determined by measuring the left hand side of
\begin{equation}
\left<0\right|\sum_x P^{r \times t}(x)\left|0\right> \propto e^{-V(r)t}
\end{equation}
that is measuring
\begin{equation}
V(r)=-\frac1t \log \left( \frac1N \sum_{U,x} Re Tr P^{r\times t}(x) \right)
\end{equation}

Fig.~\ref{staticpotential} shows the result of this computation\cite{Duncan:1998gq}. The open squares represent the static potantial for a pure $SU(3)$ gauge field theory in 4D ($s=4$) referred to as quenched QCD. 
Notice that, for long distance, $V(r)\simeq \sigma r$ where $\sigma$ is the string tension. The solid points in figure represent the same static potential with a gauge field coupled to two dynamical light quarks for two different values of the quark mass, $m$. We expect a change of regime when $V(x) > 2m$. In fact, according to current models of confinement\cite{confinement}, when the static potential exceeds the energy required to create a new couple of quark and antiquark, these new particles pop up from the vacuum and screen the interaction between the original static quarks. The plot shows an indication of this phenomenon.

\section{Fermions}

Fermions, here identified by $\psi$, are solutions of the Dirac equation. The latter can be derived from the continuum Euclidean action
\begin{equation}
\mathcal{S}^{Dirac}_E=\int \textrm{d}^4x \bar\psi^i_\alpha(x)(\gamma_{\alpha\beta}^\mu D^{ij}_\mu +m\delta_{\alpha\beta})\psi^j_\beta(x)
\label{dirac}
\end{equation}
where $D_\mu=\partial_\mu+igA_\mu(x)$ is the covariant derivative, $\alpha\beta$ are spin indices and $ij$ are gauge indices. 

From now on the gauge indices $ij$ will often be omitted and we will restrict to 4D, $s=4$, since fermions are only defined for even number of dimensions.

Notice that because of the Wick rotation, gamma matrices have to be rotated too. The Euclidean gamma matrices are hermitian ($\gamma_\mu^\dagger=\gamma_\mu$) and satisfy $\{\gamma_\mu,\gamma_\nu\}=2\delta_{\mu\nu}$. Two possible choices for Euclidean gamma matrices in four dimensions are listed in the appendix.

In the absence of gauge interaction, $D_\mu=\partial_\mu$ and the simplest symmetrical discretized derivative looks like
\begin{equation}
\partial_\mu \psi(x) \stackrel{def}{=} \frac1{2}\left[ \psi(x+\hat\mu)-\psi(x-\hat\mu) \right]
\label{partial}
\end{equation}
The most local generalization of eq.~(\ref{partial}) that preserves the gauge invariance of the action is
\begin{equation}
D_\mu \psi(x) \stackrel{def}{=} \frac1{2}\left[ U(x,\mu)\psi(x+\hat\mu)-U^\dagger(x-\mu,\mu)\psi(x-\hat\mu) \right]
\label{dcov}
\end{equation}

With this definition for the derivative, eq.~(\ref{dirac}) becomes
\begin{equation}
\mathcal{S}^{Dirac}_E=\sum_{x,x'} \bar\psi_\alpha(x)Q_{\alpha\beta}(x,x') \psi_\beta(x)
\label{latticedirac}
\end{equation}
and $Q$ is called fermionic matrix
\begin{equation}
Q^{naive}_{\alpha\beta}(x,x') = m \delta_{x,x'}\delta_{\alpha\beta} + \sum_{\mu} \gamma_{\alpha\beta}^\mu \frac1{2}\left[ U(x,\mu)\delta_{x+\hat\mu,x'}-U(x,-\mu)\delta_{x-\hat\mu,x'} \right]
\label{qnaive}
\end{equation}

There is still a problem with this naive action. 
The quark propagator S is defined as the inverse of the fermionic matrix $Q$, i.e.
\begin{equation}
S^{ij}_{\alpha\beta}(x,x')\stackrel{def}{=}\left<0\right|T\{\psi^j_\beta(x'),\bar\psi^i_\alpha(x)\}\left|0\right> = (Q^{-1})^{ij}_{\alpha\beta}(x,x')
\end{equation}

In absence of gauge field ($U(x,\mu)=1$) for a massless quark ($m=0$), for a single Fourier component of the fermionic field ($\psi(x) = e^{ip_\mu x^\mu}$) the inverse propagator reads
\begin{equation}
S^{-1}=Q=m + i\sum_\mu \gamma^\mu \sin(p_\mu)
\label{naiveprop}
\end{equation}
($p_\mu$ is in units of $1/a$ here). This propagator has $16$ poles as opposed to the single pole at $p=0$ for the continuum propagator. The additional poles arise when the spatial components of $p$ are equal to $\pi$.

The physical interpretation of the additional poles is that this naive discretization of the action describes 16 degenerate fermions as opposed to a single one. Different solutions to this problem lead to different implementations of lattice fermions. We consider here Wilson, clover, staggered, overlap, and domain wall fermions. 

\subsection{Wilson Fermions}

Wilson proposed to remove the additional poles by giving mass to the corresponding modes\cite{Wilson:1974}. This is done by adding a new term to the Lagrangian proportional to
\begin{equation}
r \bar\psi_\alpha(x)D_\mu D^\mu \psi_\alpha(x)
\end{equation}
This corresponds to replacing eq.~(\ref{qnaive}) with
\begin{eqnarray}
Q_{\alpha\beta}(x,x') &= (m+4r) \delta_{x,x'}\delta_{\alpha\beta} - \frac12 \sum_{\mu}[&(r-\gamma^\mu)_{\alpha\beta} U(x,\mu)\delta_{x+\hat\mu,x'}- \nonumber \\
&&(r+\gamma^\mu)_{\alpha\beta} U(x,-\mu)\delta_{x-\hat\mu,x'} ]
\label{qdef2}
\end{eqnarray}
Introducing the definition 
\begin{equation}
\kappa=\frac1{2m+8r}
\label{kappa}
\end{equation}
and scaling the field $\psi$, eq.~(\ref{qdef2}) can be rewritten as (omitting spin indices)
\begin{equation}
Q^{W}(x,x') = \delta_{xy} - \kappa \sum_{\mu} \left[ (r-\gamma^\mu) U(x,\mu)\delta_{x+\hat\mu,x}-(r+\gamma^\mu) U(x,-\mu)\delta_{x-\hat\mu,x'} \right]
\label{qdef3}
\end{equation}
which is the standard form for the Wilson fermionic matrix. Notice that any value of $r>0$ will do the job and one usually chooses $r\equiv 1$. The choice $r=0$ corresponds to the naive action eq.~(\ref{qnaive}).

In the Wilson fermionic action the fermion mass is traded in for the adimensional parameter $\kappa$ defined in the asymptotically free limit by eq.~(\ref{kappa}). In presence of gauge interaction $\kappa$ is renormalized and eq.~(\ref{kappa}) holds only approximatively. This renormalizion shifts the value of $\kappa$ that corresponds to chiral fermions ($m=0$). From now on we will identify with $\kappa^\ast$ that value of $\kappa$ that corresponds to a chiral fermion. $\kappa^\ast$ is not known a priori but it can be determined numerically as it corresponds to a pole of $Q^{-1}$.

For Wilson fermions on a cold configurations, $U\equiv 1$, the fermion propagator reads
\begin{equation}
S^{-1}=Q=m+i\sum_\mu \gamma^\mu \sin(p_\mu) +2r \sum_\mu \sin^2(p_\mu/2)
\end{equation}

The main problem with Wilson fermions is that the fermionic matrix for $m=0$ does not anticommute with $\gamma_5$ and therefore the action is not invariant under the global chiral symmetry 
\begin{equation}
\psi(x)\rightarrow e^{i\gamma_5\theta}\psi(x) \hskip 2cm
\bar \psi(x)\rightarrow \bar \psi(x) e^{i\gamma_5\theta}
\end{equation}
 which is a symmetry of the continuum Dirac action.

The breaking of chiral symmetry invalidates chiral perturbation theory which is used to guide guide the extrapolation of the spectrum to the limit $m \rightarrow 0$. This extrapolation is crucial in order to compute correlations involving fermions with mass smaller than $1/a$.

\subsection{Clover Fermions}

The simplest way to Symanzik improve Wilson fermions is by shifting fermions according to
\begin{equation}
\psi(x) \rightarrow (1 + c\gamma^\mu D_\mu)\psi(x)
\label{shift_field}
\end{equation}
and tune $c$ in order to cancel any $O(a)$ dependence in the correlations. The effect on the action is the same as replacing\cite{sw,luscher98}
\begin{equation}
Q^{SW}(x,y) = Q^{W}(x,y) - \frac{i r c_{SW}}{4} \sum_{\mu\neq\nu} \gamma_\mu \gamma_\nu F_{\mu\nu}(x) 
\label{qdefsw}
\end{equation}
where $F_{\mu\nu}=[D_\mu,D_\nu]$ is a lattice version of the electromagnetic tensor which, in terms of the links, can be expressed as
\begin{eqnarray}
F_{\mu\nu}&=&(B-B^\dagger)/8 \nonumber \\
B&=&P(x,\mu,\nu)+P(x,\mu,-\nu)+P(x,-\mu,\nu)+P(x,-\mu,-\nu)  \label{fmunu}
\end{eqnarray}
Eq.~(\ref{latticedirac}) with ~(\ref{qdefsw}) is referred to as Sheikoleslami-Wohlert (SW) action or simply {\it clover} action because of the expression for $F$, eq.~(\ref{fmunu}).

The value of the coefficient $c_{SW}$ does not affect the results in the continuum limit $a\rightarrow 0$ but does affect the rate of convergence in this limit and the symmetry restoration for those symmetries broken by the lattice. 

There are three standard techniques to choose the parameter $c_{SW}$

\begin{itemize}

\item  1-loop improvement\cite{Luscher} 
\begin{equation}
c_{SW}=1+1.5954/\beta
\end{equation}

\item  Tadpole improvement\cite{lepage93}
(i.e. resumming the
contribution of all tadpole 
graphs to the renormalization of the tree-level $c_{SW}$).
\begin{equation}
c_{SW}=\frac{1}{u_0^3}
\end{equation}
where $u_0$ is the average of $\frac13 Re Tr P(x,\mu,\nu)$ and it can be extracted from numerical simulations.

\item 
Non-perturbative improvement\cite{Luscher}. This is the most sophisticated 
technique. The idea behind it is that of determining the
improvement coefficients for the different operators (including $c_{SW}$) by
measuring independently on lattice the left and right-hand side of
Ward identities and imposing the constraint that
they coincide. The results for $c_{SW}$ can be
summarized by the following fitting function (valid only for $\beta>5.7$)
\begin{equation}
c_{SW}=\frac{\beta^3-3.648\beta^2-7.254\beta+6.642}{\beta^3-5.2458\beta^2}
\end{equation}
\end{itemize}

Even if different techniques give different results for
$c_{SW}$ they are consistent with each other provided the operators
are improved by adopting the same procedure. Whichever
improvement technique is used, the SW action generates correlation amplitudes
that converge to the continuum limit up to correction of the second
order in $a$ and order 1 in $g^2$ (for 1-loop) or exactly 
(for non-perturbative improvement).

\subsection{Heavy Fermions and Fermilab Action}

As mentioned previously the lattice regularization acts as infrared cut-off and prevent particles with mass $m_Q$ higher then $1/a$ to propagate properly. This presents a problem for the simulation of heavy quarks since in typical LQCD computations the lattice spacing is of the order of $(2\textrm{GeV})^{-1}$. 
 There are four ways to implement heavy fermions on the lattice.

{\bf Extrapolation}:
Perform simulations with fermion masses lighter than the cut-off, $m<1/a$, and extrapolate at the physical heavy fermion mass $m\rightarrow \bar m_Q$.

{\bf Static fermions}:
Perform simulations in the static limit, $m_Q \rightarrow \infty$. In this limit the Wilson action becomes the HQET\cite{Wise:1993ph} action with fermionic matrix given by
\begin{equation}
Q^{HQET}(x,x') = \frac{1-\gamma^0}{2} U(x,0)\delta_{x+\hat0,x}+\frac{1+\gamma^0}{2} U(x,-0)\delta_{x-\hat0,x'}
\label{qdefhq}
\end{equation}
In this limit the a fermion propagator is known exactly and it is the product of links in the time direction
\begin{eqnarray}
S(x,x') &=& \frac{1+\gamma^0}{2} U(x,0)U(x+\hat 0,0)...U(x'-\hat 0,0) \delta_{\mathbf{x},\mathbf{x'}} \theta(x_0-x'_0) +
\nonumber \\
&& \frac{1-\gamma^0}{2} (U(x,0)U(x+\hat 0,0)...U(x'-\hat 0,0))^\dagger \delta_{\mathbf{x},\mathbf{x'}} \theta(x'_0-x_0)
\end{eqnarray}
Notice how static fermions require only one spin component because no term in the action mixes different spin components.
Computations in the static limit are also useful to guide the extrapolation in the approach previously mentioned.

{\bf Non relativistic limit}:
Adopt a non-relativistic approach (NRQCD) and perform an expansion of the fermionic action around the on-shell momentum\cite{manohar} of propagating fermions $p_\mu=p_\mu^{\textrm{on-shell}}+k_\mu$. Non-relativistically $p_\mu^{\textrm{on-shell}}=m_Q v_\mu$ where $v_\mu$ is the on-shell velocity of the heavy fermion and $k_\mu$ is the off-shell momentum. In QCD, $k_\mu$ is of the order $\Lambda_{QCD}$. This expansion results in the NRQCD action described by the fermionic matrix
\begin{equation}
Q^{NRQCD}(x,x') = i \gamma^0 (D_0 + {\mathbf v} \cdot {\mathbf D} + O(1/m_Q))
\label{qdefnrqcd}
\end{equation}
NRQCD fermions require two spin components which are mixed by $O(q/m_Q^2)$ terms.
Corrections can be taken into account systematically.

{\bf Fermilab action}:
The Fermilab action is an effective action that interpolates smoothly between the regular fermions and the static limit, and eliminates errors proportional to $(a m_Q)^n$. This is achieved by taking Wilson fermions with the clover action and introducing different couplings for space-like and time-like interaction terms in the Lagrangian. Moreover, in the spirit of Symanzik, higher order operators are added to the Lagrangian to cancel discretization terms that become sizable for large masses. A Discussion of these operators up to dimension 4 in $\Lambda_{QCD}/m_Q$ (HQET) and dimension 8 in $v^\mu \gamma_\mu$ (NRQCD) can be found in \cite{andreas,Oktay:2003gk}. 

\subsection{Staggered Fermions}

The Kogut-Susskind fermions, also known as staggered fermions 
provide an alternative to Wilson fermions.

\begin{figure}
\begin{center}
\vskip -3cm
\begin{turn}{270}	
\includegraphics[height=12cm]{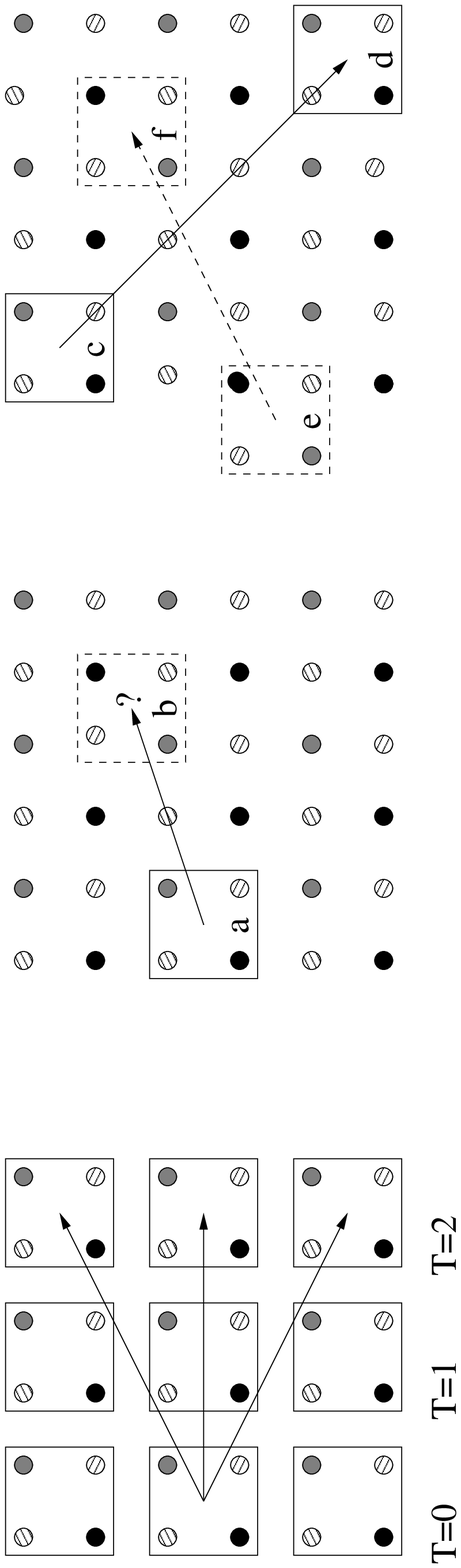}
\end{turn}
\vskip -3cm
\end{center}
\caption{Graphical representation of a 2D lattice and its blocked lattice (left). The center figure represents an example of an incorrect 2-point correlation amplitude between staggered fermions. The right figures represents two correct 2-point correlation amplitudes corresponding to different blockings of the same lattice.\label{fig:boxes}}
\end{figure}

The idea of Kogut and Susskind is that of interpreting the 16 poles of the naive fermion propagator, eq.~(\ref{naiveprop}), as due to the 4 spin components of 4 different types of fermions, here referred as {\it flavors}, which live on a blocked lattice\cite{kogut,banks,susskind,Kilcup:1987dg,Golterman:1986dz} as shown in fig.~\ref{fig:boxes}.

In order to introduce staggered fermions we will adopt the following notation:
\begin{itemize}

\item 
$x$, $x'$ will indicate points on the full lattice.

\item
$y$, $y'$ will indicate points on the blocked lattice

\item
$z$ will label the vertices of a hypercube of side $a$ so that each $x$ has a unique representation as $y+z$. $z_\mu \in {0,1}$ are the four-spacetime components of $z$.

\item
$\psi_{a\alpha}(y)$ will represent the four fermion flavors where $a$ is the flavor index and $\alpha$ is the spin index. Note that $\psi$ is now defined on the blocked lattice.

\item
$\chi(x)$ will indicate the proper staggered field which corresponds to $\psi$ but is defined on the full lattice.

\item 
We will omit  the color index since the formalism in completely transparent to it.
\end{itemize}

The map between naive lattice fermions and staggered fermions is realized by
\begin{equation}
\psi_{a\alpha}(y)=\frac12 \sum_z \Omega(y+z)_{a\alpha} \mathcal{P}(y,y+z)\chi(y+z)
\label{staggeredmap}
\end{equation}
where
\begin{equation}
\Omega(x)\stackrel{def}{=}\gamma_0^{x_0}\gamma_1^{x_1}\gamma_2^{x_2}\gamma_3^{x_3} 
\label{staggeredomega}
\end{equation}
and $\mathcal{P}(y,y+z)$ is a product of links connecting $y$ to $y+z$ that makes $\psi(y)$ transform correcly under local gauge transformations. The map in eq.~(\ref{staggeredmap}) between $\psi \leftrightarrow \chi$ is invertible.

The naive action of eqs.~(\ref{latticedirac},\ref{qnaive}) after substituting in eq.~(\ref{staggeredmap}) becomes
\begin{eqnarray}
\mathcal{S}_E&=&\sum_y \bar \psi_{a\alpha}(y) (\gamma^{\alpha\beta}_\mu D_\mu +m) \psi_{a\beta}(y) \nonumber \\
&=&\sum_{y,z,z'} \chi^\dagger(y+z) \Omega^\dagger(y+z) \mathcal{P} (\gamma_\mu D_\mu +m) \mathcal{P} \Omega(y+z') \chi(y+z') \nonumber \\
&=& \sum_{x,\mu} \frac{\eta(x,\mu)}{2}[\chi^\dagger(x)U(x,\mu)\chi(x+\mu)-\chi^\dagger(x)U(x,-\mu)\chi(x-\mu)] + \nonumber \\
&& \sum_{x} m |\chi(x)|^2 \nonumber \\
&=& \sum_{x,x'} \chi^\dagger(x)Q^{KS}(x,x')\chi(x')
\label{ksaction}
\end{eqnarray}
where
\begin{equation}
\eta(x,\mu) = \frac14 \text{tr}\left\{
\Omega ^{\dagger }(x)\gamma_\mu 
\Omega (x\pm\mu)
\right\}
= (-1)^{\sum_{\nu<\mu} x_\nu}  \label{eta}
\end{equation}
and
\begin{equation}
Q^{KS}(x,x')=m \delta_{x,x'} + \frac12 \sum_{\mu} \eta(x,\mu) [ U(x,\mu)\delta_{x+1,x'} - U(x,-\mu)\delta_{x-1,x'} ] \label{qks}
\end{equation}

Notice that in deriving the staggered action one uses a derivative term $D_\mu$ defined on the full lattice and not on the blocked lattice. This has the effect  of coupling the different fermions in nontrivial ways and breaks 
the continuum $SU(4)_{flavor}\times SU(4)_{spin}$ 
symmetry down to the discrete subgroup $SW_4 \times \Gamma_4$
where $SW_4$ is the hypercubic subgroup of Euclidean rotations, $SO(4)$, and
$\Gamma_4$ is the Clifford algebra generated by the $\gamma$ matrices.
This symmetry breaking allows us to identify the different flavors.

In fact, in the naive discretization of the 
action~(\ref{dirac}) the doubling problem is related to the lattice
symmetry\cite{Lepage:1999vj}
\begin{equation}
\psi(x) \rightarrow \tilde \psi(x) = 
\prod_\mu (i \gamma_5 \gamma_\mu)^{z_\mu} e^{i x \cdot z \pi} \psi(x)
\label{lepage_map}
\end{equation}
where  $z$ is in the hypercube.
Eq.~(\ref{lepage_map}) suggests that the naively discretized eq.~(\ref{dirac})
contains interaction terms that couple one fermion mode into another 
by emission of a hard gluon. For staggered fermions these modes correspond
to different flavors and, therefore, can be distinguished.

While Wilson fermions completely break the axial symmetry, 
staggered fermions break $SU(4)_V\times SU(4)_A$ only partially and
the subgroup $U(1)_V\times U(1)_A$ remains unbroken
thus preserving some form of the  PCAC relations\cite{sharpe} which 
are important to guide the extrapolation to the chiral limit 
of quantities of phenomenological interest involving light fermions.

Another advantage of staggered fermions over Wilson fermions is that 
the inversion of the corresponding fermionic matrix $Q$, 
i.e. the computation of  the fermion propagator, 
is about eight times faster for the same lattice size.

One disadvantage of staggered fermions is that they entangle spacetime and flavor symmetries thus 
making it difficult to engineer physical states of definite quantum numbers. 

In terms of the spinors the most general pseudoscalar meson can be written as 
\begin{equation}
\pi _{\xi }(y)=\xi ^{ab}\overline{\psi }_{\alpha ,a}(y)\Gamma _{\alpha \beta
}\psi _{\beta ,b}(y) 
\label{pionop}
\end{equation}
where $\Gamma$ is $\gamma^5$ or $\gamma^0 \gamma^5$.

For this operator to have the quantum numbers 
of a pion, $\xi ^{ab}$ must be an element of the 
{\bf 15} representation of $SU(4)_{flavor}$. The choice $\xi ^{ab}=\delta_{ab}$ would correspond to the singlet, the $\eta_1$.

Different choices of basis for the $\xi ^{ab}$ matrices are present 
in the literature and they are
equivalent to each other up to an unitary transformation since there is only
one irreducible representation of the Clifford algebra in dimension 4, the
algebra of ordinary gamma matrices (associated to the group $\Gamma _{4}$).
Therefore, according with usual conventions, we choose the following basis
for the $\xi $ matrices
\begin{equation}
\xi _{(5)}=(\gamma ^{5})^{*},\qquad \xi _{(\mu )}=(\gamma ^{\mu
})^{*},\qquad \xi _{(\mu 5)}=(\gamma ^{\mu }\gamma ^{5})^{*},\qquad \xi
_{(\mu \nu )}=(\gamma ^{\mu }\gamma ^{\nu })^{*} 
\end{equation}

Fig.~\ref{fig:boxes} represent a lattice, a blocked lattice, and how staggered mesons may propagate from one hypercube to another hypercube. Notice that source and destination hypercube has to be consistent with the same blocking of the lattice or the meson propagator will correspond to a nontrivial mix of the different mesons and will exhibit an oscillating behavior.

Staggered fermions with the action eq.~(\ref{ksaction}) converge quadratically to the continuum. One way to Symanzik improve this action and achieve $O(a^2)$ improvement is by replacing each link that appears in eq.~(\ref{qks}) with a weighted sum of links and paths connecting the same endpoints as the link. The choice of the weigth factors that cancels all $O(a^2)$ effects is called ASQTAD action\cite{asqtad}.

\subsection{Chiral Fermions: Overlap}

In continuum space the massless fermionic matrix is $Q=\sum_\mu \gamma^\mu (\partial_\mu + igA_\mu)$ and it satisfies the chirality condition
\begin{equation}
Q \gamma^5 + \gamma^5 Q =0 
\label{chiral1}
\end{equation}
therefore eigenstates of $P_L=(1-\gamma^5)/2$ and $P_R=(1+\gamma^5)/2$ are not mixed by the action. The only term in the action that couples $L$ and $R$ chirality states is the mass term.

Eq.~(\ref{chiral1}) does not hold for $Q^{W}$, $Q^{SW}$ and $Q^{KS}$ and, in fact, the Nielsen-Ninomiya theorem~\cite{Nielsen} states that it is not possible to define a local, translationally invariant, hermitian $Q$ that preserves chiral symmetry and does not cause doublers. Nevertheless one may look for a fermion formulation on the lattice that exhibit chiral symmetry and no doublers by requiring that chiral symmetry holds for on-shell states only.
For on shell states chiral transformation can be written as
\begin{equation}
\psi \rightarrow e^{i\theta \gamma^5 (1-a Q/2)} \psi \hskip 2cm
\bar \psi \rightarrow \bar \psi e^{i\theta \gamma^5 (1-a Q/2)} 
\label{onshellchiral}
\end{equation}
In fact on shell $Q \psi=0$ and eq.~(\ref{onshellchiral}) becomes the prdinary chiral symmetry transormation.
Requiring that the fermionic matrix $Q$ be invariant under infinitesimal transformation leads to the Ginsparg-Wilson relation
\begin{equation}
Q \gamma^5 + \gamma^5 Q =a Q \gamma^5 Q
\label{ginspargwilson}
\end{equation}

Neuberger\cite{Neuberger1998my,neuberger} proposed the first fermionic matrix $Q$ that satisfies the Ginsparg-Wilson relation and thus provides exact chiral symmetry on the lattice
\begin{equation}
Q^{Overlap}=1+\gamma^5 \text{sign}[\gamma^5(Q^{W}-1)]
\label{overlap}
\end{equation}
This formulation is referred to as overlap fermions.  

Since any hermitian matrix $\Sigma$ can be written as $\Sigma = \Lambda D \Lambda^\dagger$ where $D$ is a diagonal matrix, one can define any function $f$ of a hermitian matrix via
\begin{equation}
f(\Sigma) = \Lambda f(D) \Lambda^\dagger
\end{equation}
and $f(D)$ is a diagonal matrix with diagonal elements given by $f(D)_{ii}=f(D_{ii})$. The argument of sign function in the definition of the overlap fermionic matrix is hermitian therefore eq.~(\ref{overlap}) is well defined.

The sign function can be approximated by polynomials and, in general, it is more difficult to deal with than Wilson, clover or even staggered fermions. All numerical techniques\cite{Blum2000kn} for calculating a $\det Q^{Overlap}$ as required for dynamical fermions and for calculating $(Q^{Overlap})^{-1}$ can be viewed as different ways of approximating eq.~(\ref{overlap}).

\subsection{Chiral Femions: Domain Wall}

An alternative way to implement lattice chiral fermions was developed by Kaplan\cite{kaplan} in 1992. It is referred to as using Domain Wall (DW) fermions\cite{shamir1,shamir2,Furman1994ky,Blum2000kn}. This formulation consists of replacing each fermion $\psi_\alpha(x)$ with multiple fermions labeled by a new index $k$
\begin{equation}
\psi_\alpha(x) \rightarrow \Psi_{\alpha,k}(x)
\end{equation}
The DW action is
\begin{equation}
\mathcal{S}_E^{DW}=\sum_{x,k} \bar \Psi_{\alpha, k}(x) Q^{DW}_{\alpha\beta,kk'}(x,x') \Psi_{\beta,k'}(x')
\end{equation}
where
\begin{eqnarray}
Q^{DW}_{\alpha\beta,kk'}(x,x')&=&Q^{W}_{\alpha\beta}(x,x')\delta_{kk'} + \nonumber \\
&& (P_L)_{\alpha\beta} \theta(k<N_5-1) \delta_{k+1,k'} + (P_R)_{\alpha\beta} \theta(k>0) \delta_{k-1,k'} + \nonumber \\
&& (P_L)_{\alpha\beta} \delta_{k,N_5-1}\delta_{k',0} + (P_R)_{\alpha\beta} \delta_{k,0}\delta_{k',N_5-1} 
\label{qdw}
\end{eqnarray}
$k_5$ and $N_5$ are parameters of the model. The index $k$ runs from $0$ to $N_5-1$ and it is usually interpreted as a 5th dimension of the system. Notice that the gauge field that appears in $Q^{W}$ is independent on this 5th dimension.

The computation of $\psi'=Q^{-1}\psi$ for a regular 4 dimensional input fermionic field $\psi$ is performed in three steps:

\begin{itemize}

\item
The input field $\psi$ is mapped into the 5-dimensional DW field $\Psi$. $L$ components are mapped into the $k=0$ slice and $R$ components are mapped into the $k=N_5-1$ slice.
\begin{eqnarray}
\Psi_{\alpha,0} (x)&=&(P_L)_{\alpha\beta} \psi_{\beta}(x) \nonumber \\
\Psi_{\alpha,k} (x)&=&0 \ \ \text{ for }\ \ 0<k<N_5-1  \nonumber \\
\Psi_{\alpha,N_5-1}&=&(P_R)_{\alpha\beta} \psi_{\beta}(x) \\
\end{eqnarray}

\item 
The DW fermionic matrix, eq.~(\ref{qdw}), is inverted numerically using one of the algorithms explained later
\begin{equation}
\Psi'=(Q^{DW})^{-1}\Psi
\end{equation}

\item
The output DW field $\Psi'$ is mapped back into the output 4 dimensional field $\psi'$ by
\begin{equation}
\psi'_\alpha(x) = (P_L)_{\alpha\beta} \Psi'_{\beta,0}(x) + (P_R)_{\alpha\beta} \Psi'_{\beta,N_5-1}(x)
\end{equation}

\end{itemize}

The effect of the DW action is that of projecting the $L$ modes of the 4D fermion to one wall and the $R$ modes to the other wall, thus different chiralities are mixed only by the explicit mass term in $Q^{DW}$, eq.~(\ref{qdw}). The use of the Wilson action in each slice of the extra dimension $k$ guarantees that no doublers are present.

It is possible to translate the DW fermions into an equivalent four-dimensional approximation for the Neuberger operator\cite{Borici1999zw}, eq.~(\ref{overlap}) and therefore $Q^{Overlap}$ is a local operator. In fact, any rational polynomial approximation of the overlap operator is equivalent to a domain wall formulation with a finite 5th dimension~\cite{kennedy:2004}. 

\subsection{Quenched and Dynamical Fermions}

Including fermionic field variables in the MCMC configurations is not practical. The usual technique to deal with fermions is integrate them out analytically so that fermions do not appear in the PI measure

Let's consider a system consisting of a gauge field coupled to $n_f$ degenerate fermions
\begin{eqnarray}
\mathcal{S}_E&=&\mathcal{S}^{gauge}_E+n_f \mathcal{S}^{fermi}_E \nonumber \\
&=&\frac{-\beta}{2n}\sum_{x,\mu\neq\nu} Re Tr P(x,\mu,\nu) + n_f \sum_{x,y} \psi_\alpha(x) Q_{\alpha\beta}(x,y) \psi_\beta(y)
\label{fullqcd1}
\end{eqnarray}

For a typical correlation amplitude, fermions can be integrated out as follows
\begin{eqnarray}
\left<0\right|... \psi_\alpha(x)\bar\psi_\beta(x')...\left|0\right> &=& 
\int [\textrm{d}U][\textrm{d}\psi]... \psi_\alpha(x)\bar\psi_\beta(x')... e^{-\mathcal{S}_E^{gauge}-n_f \mathcal{S}_E^{fermi}} \\
&=&
\int [\textrm{d}U]... Q^{-1}_{\alpha\beta}(x,x')... e^{-\mathcal{S}_E^{gauge}+n_f \log\det Q} \\
&=&
\frac1N \sum_{U} ...Q^{-1}_{\alpha\beta}(x,x')... 
\end{eqnarray}
where the field configurations are now generated at random by sampling from a probabily distribution proportional to $\exp(-\mathcal{S}^{full}_E)$ with
\begin{equation}
\mathcal{S}^{full}_E=\mathcal{S}_E^{gauge}-n_f \log\det Q
\label{fullaction}
\end{equation}

In case the correlation amplitudes involve multiple possible Wick contractions, one has to sum over all possible Wick contractions.
\begin{eqnarray}
\left<0\right|... \psi_\alpha(x)\bar\psi_\beta(x')...\psi_\gamma(x'')\bar\psi_\delta(x''')\left|0\right> 
&=& 
\frac1N \sum_{U} ...Q^{-1}_{\alpha\beta}(x,x')...Q^{-1}_{\gamma\delta}(x'',x''')... + \nonumber \\
&&\phantom{\frac1N \sum_{U}} ...Q^{-1}_{\alpha\delta}(x,x''')...Q^{-1}_{\gamma\beta}(x'',x')... 
\end{eqnarray}

For staggered fermions, since $Q^{KS}$ represents four degenerate flavors as opposed to a single one, therefore eq.~(\ref{fullaction}) must be replaced by
\begin{equation}
\mathcal{S}^{full KS}_E=\mathcal{S}_E^{gauge}-\frac{n_f}{4} \log\det Q^{KS}
\label{fullactionks}
\end{equation}

It is still debated whether the above procedure is correct, since there is no known local operator that corresponds to the 4th root of $Q^{KS}$. Anyway, there are indications from numerical studies\cite{Durr:2004ta} in 2D that the following relation may hold in 4D
\begin{equation}
(\det Q^{KS})^{1/4} = \det Q^{Overlap} + O(a^2)
\end{equation}
which would provide a solid theorical justification for eq.~(\ref{fullactionks}).

From the definition it is evident that $Q$ is a sparse matrix. For Wilson fermions its dimensions are $M \times M$ and $M = 8 n (L/a)^4$ (real and imaginary part $\times$ 4 spin components $\times$ $n$ color components $\times$ number of lattice sites); $Q$ has diagonal elements set to $1$ and has only 4 other elements different from zero for each row and column (corresponding to $+\mu$ and $-\mu$). For staggered fermions $Q^{KS}$ is 16 times smaller because it has no spin indices and this makes its inversion much faster. For domain wall fermions $Q^{DW}$ is $N_5^2$ larger than $Q^{W}$ thus making the inversion of domain wall fermions much slower than for Wilson fermions.

One approximation that has been used and abused consists of setting $n_f=0$ in the full action. This approximation is called {\it quenching}. It has the effect of ignoring second quantization for fermions. This is equivalent to, in a perturbative language, ignoring fermion loops. Quenching introduces unknown systematic errors in the computation and its only justification is that the computation of $\det Q$ is non-local therefore generating the Markov Chain with $n_f\neq 0$ is very computing intensive.

For some quantities such as the static quark potential, fig.~\ref{staticpotential}, the effect of quenching is very small. For other quantities such as the light spectrum of QCD, its effect is sizable, fig.~\ref{theoryvsexperiment}. Quenching also affects the behavior of the spectrum in the chiral limit as explained in \cite{Sharpe:1995qp}.

The contributions $\det Q$ to the action is referred to as {\it dynamical fermions} or {\it sea quarks} in the context of LQCD. 

\subsection{CPTH Theorem on the Lattice}

Lattice correlation amplitudes computed using the action in eq.~(\ref{fullqcd1}) are invariant under charge conjugation, C, parity, P, time reversal, T, and a new symmetry, H. These symmetries apply to the measurement of an operator on each gauge configuration $U$.

It is useful to write how these symmetries affect a fermion propagator $S=Q^{-1}$ and make explicit the dependence on all indices and on the gauge configuration $U$.

\begin{itemize}

\item {\bf Charge conjugation, $C$}: 
\begin{equation}
S_{\alpha \beta}^{ij}(x,y,U)=
(\gamma^0\gamma^2)_{\alpha\alpha'} S_{\alpha' \beta'}^{ji}(y,x,U^C)
(\gamma^2\gamma^0)_{\beta'\beta}
\end{equation}

\item {\bf Parity, $P$}:
\begin{equation}
S_{\alpha \beta}^{ij}(x,y,U)=
\gamma^0_{\alpha\alpha'} S_{\alpha' \beta'}^{ij}(x^P,y^P,U^P)
\gamma^0_{\beta'\beta} 
\end{equation}

\item {\bf Time reversal, $T$}: 
\begin{equation}
S_{\alpha \beta}^{ij}(x,y,U)=
(\gamma^0\gamma^5)_{\alpha\alpha'} S_{\alpha' \beta'}^{ij}(x^T,y^T,U^T)
(\gamma^5\gamma^0)_{\beta'\beta}
\end{equation}

\item {\bf $H$ symmetry}: 
\begin{equation}
S_{\alpha \beta}^{ij}(x,y,U)=
\gamma^5_{\beta\alpha'} S_{\alpha' \beta'}^{ji}(y,x,U)
\gamma^5_{\beta'\alpha} 
\end{equation}
\end{itemize}

$U^P,U^C,U^T$ are the parity reversed, charge conjugate, and time reversed
gauge configurations respectively. 

\subsection{Inversion Algorithms}

\begin{figure}[t]
\begin{Verbatim}[frame=single, numbers=left, numbersep=3pt, fontfamily=helvetica, commandchars=\\\{\}, codes={\catcode`$=3\catcode`^=7}]
Algorithm: Minimal Residual Inverter
Input: $\psi$, $Q$
Output: $\psi^\prime$
Temporary fields: $q$,$r$

$r=\psi-Q\psi$
$\psi^\prime=\psi$
do 
    $q=Qr$
    $\alpha=(q\cdot r)/(q \cdot q)$
    $\psi^\prime=\psi^\prime+\alpha r$
    $r=r-\alpha q$
    $residue=r \cdot r$
while $residue>precision$
\end{Verbatim}
\caption{Minimal Residual, a numerical algorithm to compute $\psi'=Q^{-1}\psi$ by minimizing $Q\psi'=\psi$.\label{MinRes}}
\end{figure}

\begin{figure}[t]
\begin{Verbatim}[frame=single, numbers=left, numbersep=3pt, fontfamily=helvetica, commandchars=\\\{\}, codes={\catcode`$=3\catcode`^=7}]
Algorithm: Stabilized Biconjugate Gradient
Input: $\psi$, $Q$
Output: $\psi^\prime$
Temporary fields: $p$,$q$,$r$,$s$,$t$

$\psi^\prime=\psi$
$r=\psi-Q\psi$
$q=r$
$p=0$ (zero field)
$s=0$ (zero field)    
$\rho=\rho^\prime=\alpha=\omega=1$
do      
    $\rho=q\cdot r$
    $\beta=(\rho/\rho^\prime)(\alpha/\omega)$
    $\rho^\prime=\rho$
    $p=r+\beta p-\beta\omega s$
    $s=Qp$
    $\alpha=\rho/(q \cdot s)$
    $r=r-\alpha s$
    $t=Qr$
    $\omega=(t \cdot r)/(t \cdot t)$
    $\psi^\prime=\psi^\prime+\omega r + \alpha p$
    $residue=r \cdot r$      
while $residue>precision$
\end{Verbatim}
\caption{Stabilized Biconjugate Gradient, another numerical algorithm to compute $\psi'=Q^{-1}\psi$ by minimizing $Q\psi'=\psi$.\label{BiCGStab}}
\end{figure}

One way to invert the matrix $Q$ is by using a stochastic technique.

In fact for any hermitian positive definite matrix $\Sigma$ the following exact relation holds:
\begin{equation}
\Sigma^{-1}_{ij} =\frac1Z \int \textrm{d}\phi_0 ...\textrm{d}\phi_{n-1} \phi^\ast_j \phi_i e^{-\phi_n^\ast \Sigma_{nm} \phi_m}
\end{equation}
And substituting in the above equation $\Sigma=Q^\dagger Q$ and multiplying both terms by $Q^\dagger$ one obtains
\begin{equation}
(Q^{-1})_{\alpha\beta}^{ij}(x,x') = \frac1Z \int [\textrm{d}\phi] (Q\phi)_\beta^{j\,\ast}(x') \phi_\alpha^i(x) e^{-\phi^\dagger (Q^\dagger Q) \phi}
\end{equation}
This multidimensional integral can be computed via Monte Carlo. The field $\phi$ in this context is usually referred to as pseudo-fermionic field. The stochastic computation of the full inverse propagator must be done for each gauge configuration and therefore it is computationally expensive. Various techniques for reducing the variance in the integration and reduce the statistical noise have been proposed by many authors\cite{michael,juge}.

In most cases one does not need the full inverse $Q^{-1}$ and it suffices to solve in $\psi'$ the equation
\begin{equation}
Q \psi' = \psi \leftrightarrow \psi' = Q^{-1} \psi
\end{equation}
for a small set of given input vectors $\psi$s.
This inversion can be performed by minimizing the norm of the residual vector
\begin{equation}
r = \psi - Q\psi'
\end{equation}
The two algorithms commonly used for performing this numerical minimization are the Minimal Residual, fig.~\ref{MinRes} and Stabilized Biconjugate Gradient fig.~\ref{BiCGStab}%
\footnote{%
In writing fermionic algorithms we adopted the following notation:
\begin{eqnarray}
\psi' \cdot \psi &\rightarrow& \sum_{x,\alpha,i} \psi^{\prime\, i\, \ast}_\alpha(x)=\psi_\alpha^i(x) \nonumber \\
|\psi|^2 &\rightarrow& \sum_{x,\alpha,i} \psi^{i\, \ast}_\alpha(x)=\psi_\alpha^i(x) \nonumber \\
\psi' = Q \psi &\rightarrow& \psi^{\prime\, i\, \ast}_\alpha(x)=\sum_{y,\beta,j} Q_{\alpha\beta}^{ij}(x,y)\psi_\beta^j(y) 
\end{eqnarray}
}

\subsection{Dynamical Fermions Algorithms}

For any Hermitian matrix $\Sigma$
\begin{equation}
\det \Sigma =\frac1Z \int \textrm{d}\phi_0 ...\textrm{d}\phi_{n-1} e^{-\phi_n^\ast (\Sigma^{-1})_{nm} \phi_m}
\end{equation}
Thus for two degenerate flavors the contribution to the action due to fermions is
\begin{equation}
(\det Q)^2 = \det Q^\dagger Q = \frac1Z \int [\textrm{d}\phi] 
e^{-\phi^\dagger (Q^\dagger Q)^{-1} \phi}
\label{detq2}
\end{equation}
where $\phi$ are pseudobosonic fields. Eq.~(\ref{detq2}) can be evaulated numerically using Monte Carlo.

Various techniques based on the above equation have been developed to speed up the MCMC and to take into account odd numbers of dynamical quarks. 

The most common MCMC algorithm for dynamical fermions is the Hybrid Monte Carlo algorithm discussed in \cite{hmc1,hmc2}.

The technique used in \cite{Duncan:1998gq} consists of observing that $\det Q^\dagger Q = (\det \gamma^5 Q)^2$ where $\gamma^5 Q$ is hermitian, and approximating the determinant with a truncated determinant, defined as the product of the eigenvalues of $\gamma^5 Q$ below some cut-off. The eigenvalues are computed by diagonalizing $\gamma^5 Q$ using the Lanczos algorithm.

One of the most promising techniques in terms of efficiency for light fermion mass is a version of the Hybrid Monte Carlo based on the Schwartz Alternating Procedure discussed in \cite{luescherhmc1} and \cite{luescherhmc2}.

Theoretical work on algorithms for dynamical overlap fermions are proposed in \cite{Neuberger:1999re} and \cite{DeGrand:2004nq}. Some preliminary phenomenological results can be found in \cite{Fodor:2004wx}. Algorithms for dynamical domain wall fermions are discussed in \cite{Chen:1998xw} and \cite{Aoki:2004ht}.

\subsection{Example: Pion Mass and Decay Constant}

\begin{figure}[t]
\begin{Verbatim}[frame=single, numbers=left, numbersep=3pt, fontfamily=helvetica, commandchars=\\\{\}, codes={\catcode`$=3\catcode`^=7}]
Algorithm: Compute the static quark-antiquark potential
Input: $\beta$, $\kappa$, size of gauge group $n$, number of MCMC steps $N$
Output: $C^\pi(t)$

Create local array $C^\pi(t)$ and initialize it to zero

for each lattice site $x$
    for each direction $\mu$
        set $U(x,\mu)$ to a random element of the gauge group $SU(n)$

for $c=1$ to $N$
    replace $U$ with the next configuration in the MCMC
    for each spin component $a$
        for each color component $i$
            make a field $\psi(x)=0$
            for each lattice site $x$ on timeslice $x^0=0$
                set $\psi^{\alpha,i}(x=0)=1$
            compute $\psi^\prime=Q^{-1}\psi$             
            for each lattice site $x$
                add $|\psi^\prime(x)|^2\text{ to }C^\pi(t=x^0)$
return $C^\pi(t)$
\end{Verbatim}
\caption{Example of algorithm to compute a pion progator. Notice the role of steps 7-9 is to create the initial configuration, step 11 loops over the Markov chain, step 12 creates the next configuration in the chain (using Metropolis, Gibbs sampling or other algorithm), steps 14-18 measure the operator, and step 19-20 average the results separately for each lattice time-slice.\label{fig:algotrithmpi}}
\end{figure}

\begin{figure}[t]
\begin{center}
\includegraphics[width=3in]{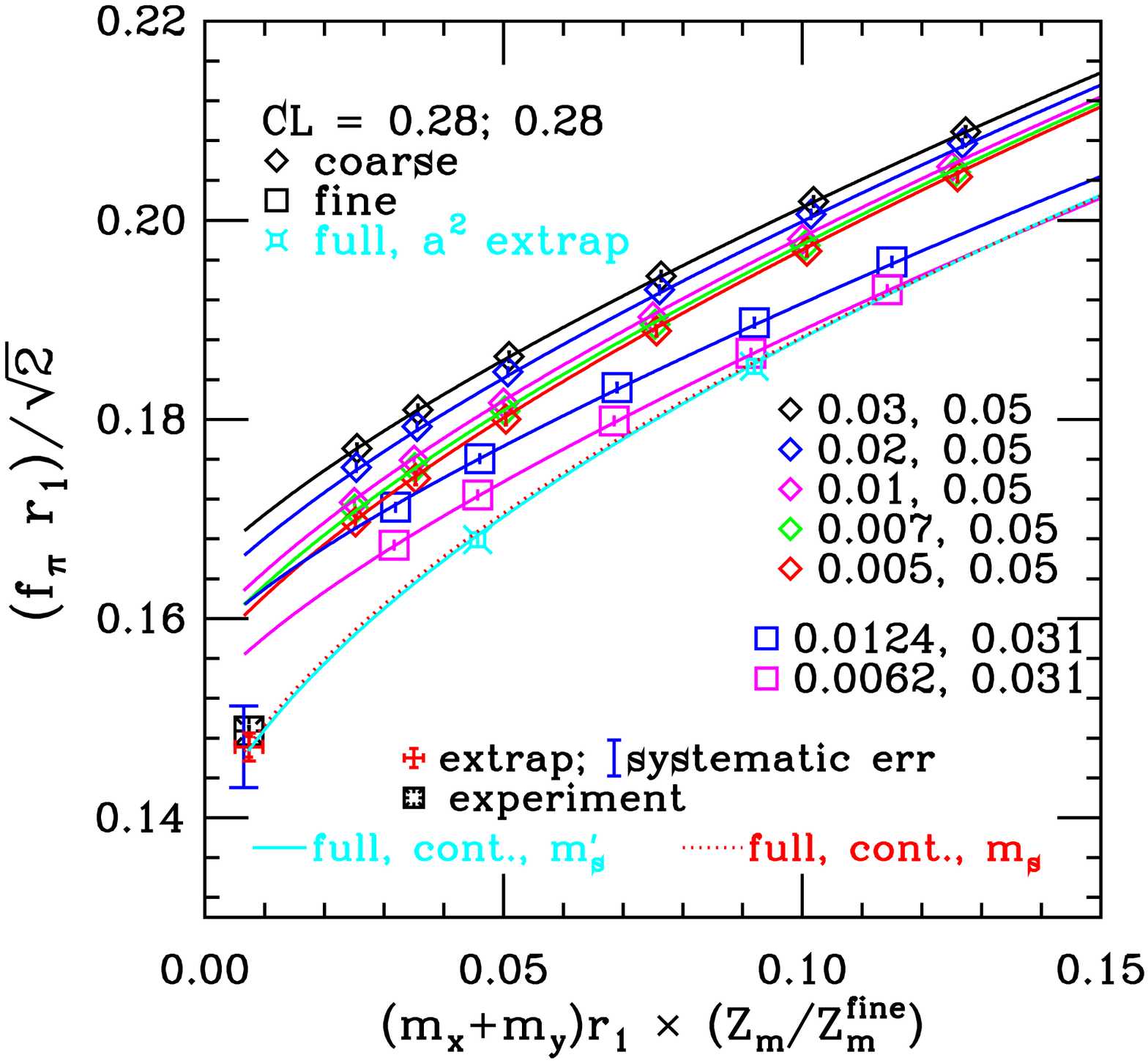}
\end{center}
\caption{Global fit of partially quenched data from the MILC collaboration$^{63}$
and chiral extrapolation of $f_\pi$ for various input values of the propagating and dynamical quarks' masses. Here $m_x$ and $m_y$ represents the bare $u$ and $d$ quark masses respectively.~\label{massextrapolation}}
\end{figure}

We consider here, as an example, the computation of the mass of the pion, $m_\pi$, and the pion decay constant, $f_\pi$,
for a $SU(n)$ gauge theory (for $n=3$ this is QCD) as defined by the action in eq.~(\ref{fullaction}).

In order to be able to put a pion on the lattice one needs to have a definition of the former, i.e. one needs to have an operator function of the gauge and quark fields that has the pion as lowest eigenstate. 

The pion is the lightest pseudo-scalar in the theory and the simplest operator that transforms as a pseudo-scalar under rotations is
\begin{equation}
\Psi_\pi(x) \stackrel{def}{=} \sum_{\alpha,\beta} \bar\psi(x)_{\alpha} \gamma^{\alpha\beta}_5 \psi_{\beta} (x)
\end{equation}
We identify with $\left|E_k\right>$ its eigenstates and $E_k$ the corresponding ordered eigenvalues so that, by definition, $E_0$ is $m_\pi$ and $\left|E_0\right>$ is $\left|\pi\right>$.
A quantum mechanical computation shows that
\begin{eqnarray}
C_\pi(y_0-x_0) &\stackrel{def}{=}&
FT^{0}_{\mathbf{x}} FT^{0}_{\mathbf{y}} \left<0\right| \Psi(x) \Psi^\dagger(y) \left| 0 \right> \\ 
&=& \sum_k FT^0_{\mathbf{x}} FT^0_{\mathbf{y}} \left<0\right| 
\Psi(x) \left| E_k \right> \frac{1}{2 E_k} \left<E_k\right| \Psi^\dagger(y) \left| 0 \right> \nonumber \\
&=& \sum_k \frac{\left| \left<0\right| \Psi(0) \left| E_k \right> \right|^2}{2 E_k} e^{i E_k x_0} e^{-i E_k y_0} \nonumber \\
&=& \frac{\left| \left<0\right| \Psi(0) \left| \pi \right> \right|^2}{2m_\pi} e^{-i m_\pi (y_0-x_0)} + ... \nonumber \\
&=& \frac{f_\pi^2 m_\pi}{2} e^{-i m_\pi (y_0-x_0)} + ... \label{cpi}
\end{eqnarray}

\begin{figure}[t]
\begin{center}
\begin{tabular*}{\hsize}{@{}@{\extracolsep{\fill}}|llll|} \hline
State & $I^G$ & $J^{PC}$ & Operator $\Psi$ \\ \hline
scalar & $1^{-}$ & $0^{++}$ & $\bar \psi \psi^{\prime }$ \\ 
& $1^{-}$ & $0^{++}$ & $\bar \psi \gamma ^0\psi^{\prime }$ \\ 
pseudoscalar & $1^{-}$ & $0^{-+}$ & $\bar \psi \gamma ^5\psi^{\prime }$ \\ 
& $1^{-}$ & $0^{-+}$ & $\bar \psi \gamma ^0\gamma ^5\psi^{\prime }$ \\ 
vector & $1^{+}$ & $1^{--}$ & $\bar \psi \gamma ^\mu \psi^{\prime }$ \\ 
& $1^{+}$ & $1^{--}$ & $\bar \psi \gamma ^\mu\gamma ^0\psi^{\prime }$ \\ 
axial & $1^{-}$ & $1^{++}$ & $\bar \psi \gamma ^\mu\gamma ^5\psi^{\prime }$ \\ 
tensor & $1^{+}$ & $1^{+-}$ & $\bar \psi \gamma ^\mu\gamma ^j\psi^{\prime }$ \\ 
octet & $\frac 12$ & $\frac 12^{-}$ & $(\psi^{Ti}\gamma ^2\gamma
^0\psi^{\prime j})(\gamma ^5\psi^{\prime \prime k }) \varepsilon _{ijk}$ \\ 
& $\frac 12$ & $\frac 12^{-}$ & $(\psi^{Ti}\gamma ^2\gamma ^0\gamma
^5\psi^{\prime j})(\psi^{\prime \prime k}) \varepsilon _{ijk}$ \\ 
decuplet & $\frac 32$ & $\frac 32^{+}$ & $(\psi^{Ti}\gamma ^2\gamma ^0\gamma
^i\psi^{\prime j})(\psi^{\prime \prime k}) \varepsilon _{ijk}$ \\ \hline
\end{tabular*}
\end{center}
\caption{Example of currents used on lattice and their relative
quantum numbers. $\psi,\psi^{\prime }$ and $\psi^{\prime \prime }$ are
different flavors. The superscripts $i$, $j$ and $k$ are gauge indices. 
\label{latticecurrents}} 
\end{figure}

\begin{figure}[t]
\begin{center}
\begin{turn}{270}	
\includegraphics[width=4in]{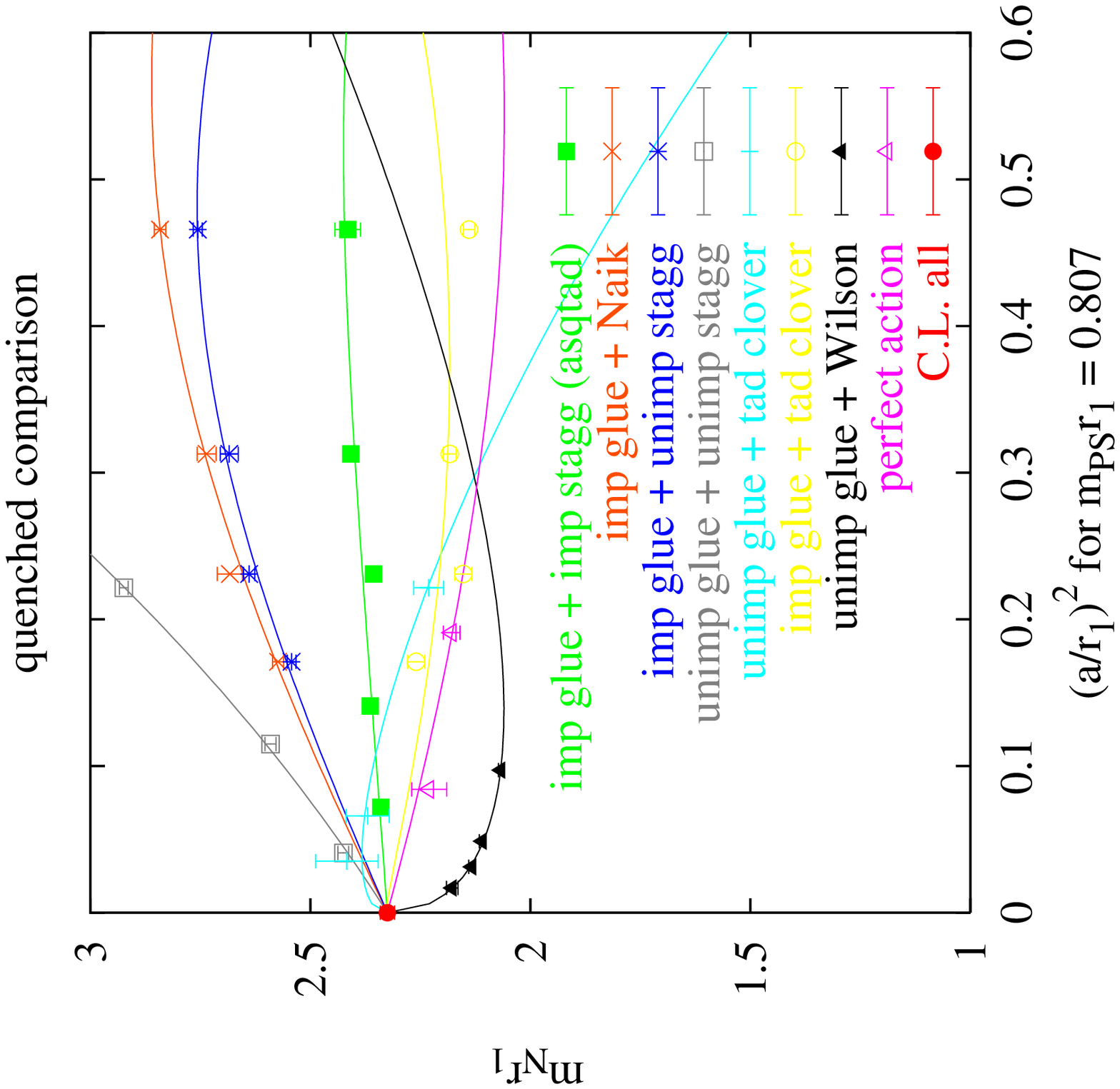}
\end{turn}	
\end{center}
\caption{The plot shows the extrapolation to the continuum limit of the neutron mass computed in LQCD using various types of fermions$^{63}$. Improved glue refers to an $O(a^2)$ improved gauge action. Different formulations agree with each other within the error but clover and staggered converge faster then Wilson as expected. $m_{PS}$ in the plot refers to the pion mass which is used to set the unit scale $r_1$.\label{variousquarks}}
\end{figure}

Here $FT^{0}_{\mathbf{x}}=\sum_{\mathbf{x}}$ is the Fourier transform in the spatial components of $x$ at zero momentum. The dots indicate exponential terms that oscillate faster that the leading term. The last step follows from the definition of $f_\pi$, the pion decay constant.

After a Wick rotation
\begin{equation}
C_\pi(y_0-x_0) = \frac{f_\pi^2 m_\pi}{2}  e^{- m_\pi (y_0-x_0)} + ...
\label{c2}
\end{equation}
and the fast oscillating terms, represented by the dots, are replaced by fast decaying exponentials. For $y_0-x_0=t$ and considering periodic boundary conditions
\begin{equation}
C_\pi(t) = \frac{f_\pi^2 m_\pi}{2} [ e^{-m_\pi (t)} + e^{-m_\pi (L-t)} ] + ... 
\label{c2_1}
\end{equation}

The lattice formulation of QCD provides the numerical technique to evaluate the left hand side of eq.~(\ref{cpi}) 
\begin{eqnarray}
C_\pi(t) &=& FT^{0}_{\mathbf{x}} FT^{0}_{\mathbf{y}} \int [\textrm{d}\phi][\textrm{d}A] \Psi^\dagger(x) \Psi(y) e^{-\mathcal{S}_E} \nonumber \\
&\simeq&\frac1N \sum_{U} FT^{0}_{\mathbf{x}} FT^{0}_{\mathbf{y}} Re Tr ( \gamma^5 Q^{-1} (x,y) \gamma^5 Q^{-1}(y,x) ) \nonumber \\
&=& \frac1N \sum_{U} \sum_{x} \sum_{y} | Q^{-1}(x,y) |^2 \delta(t-|y_0-x_0|)
\label{c2_2}
\end{eqnarray}
In the last step we used H symmetry on the second propagator.

By computing eq.~(\ref{c2_2}) for different values of $t$ and fitting the results with eq.~(\ref{c2_1}) one can extract both $m_\pi$ and $f_\pi$. Some numerical results\cite{Aubin2004fs} for $f_\pi$ computed from eq.~(\ref{c2_2}) for different values of the quark masses are shown in fig.~\ref{variousquarks}. The extrapolated $f_\pi$ is compared with the experimental results.

In practice, because of dimensional transmutation, one always computes pure numbers such as $m_{\pi}$ in units of $1/a$ and one can eliminate such dependence on $a$ by computing ratios of masses or other dimensionless ratios.

Similarly one can compute masses and decay constants of other particles by choosing the right operator. A list of operators for various interesting states is listed in fig.~\ref{latticecurrents}. For a deeper analysis on how to built this type of operators for baryons can be found in \cite{Basak:2005ir}.

Fig.~\ref{variousquarks} shows the computation of the neutron mass for different fermion formulations at different lattice spacing\cite{Davies:2004hc}. As expected, within error, they all agree with each other in the contiuum limit.

\section{Error Analysis}

\begin{figure}[t]
\begin{center}
\includegraphics[width=3in]{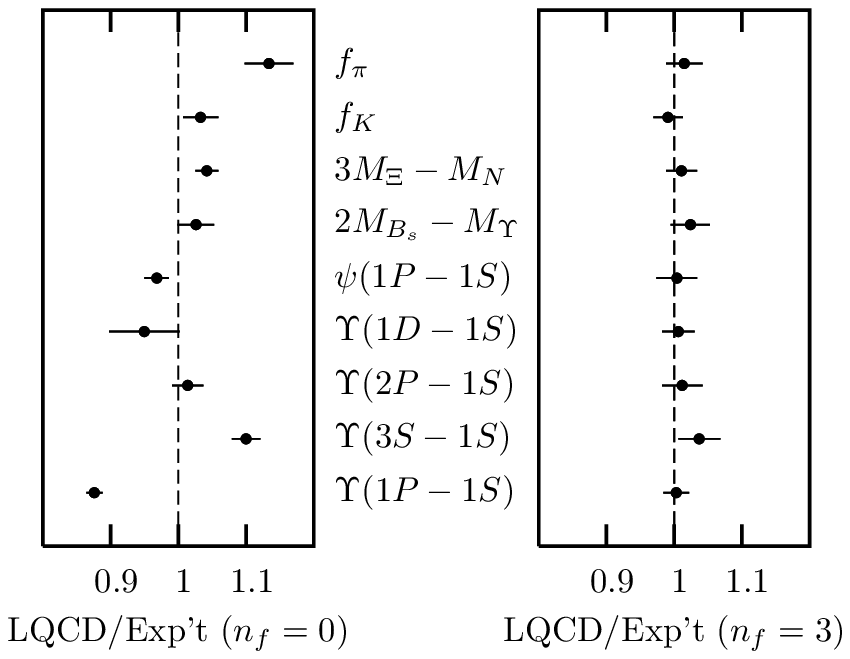}
\end{center}
\caption{Recent LQCD results from the MILC collaboration$^{68}$
. The two images show ratios of LQCD results over experimental results for various QCD quantities of phenomenological interest  without and with dynamical fermions, respectively. Dynamical fermions are implemented using $O(a^2)$ improved staggered fermions.\label{theoryvsexperiment}}
\end{figure}

\begin{figure}[t]
\begin{center}
\includegraphics[width=4in]{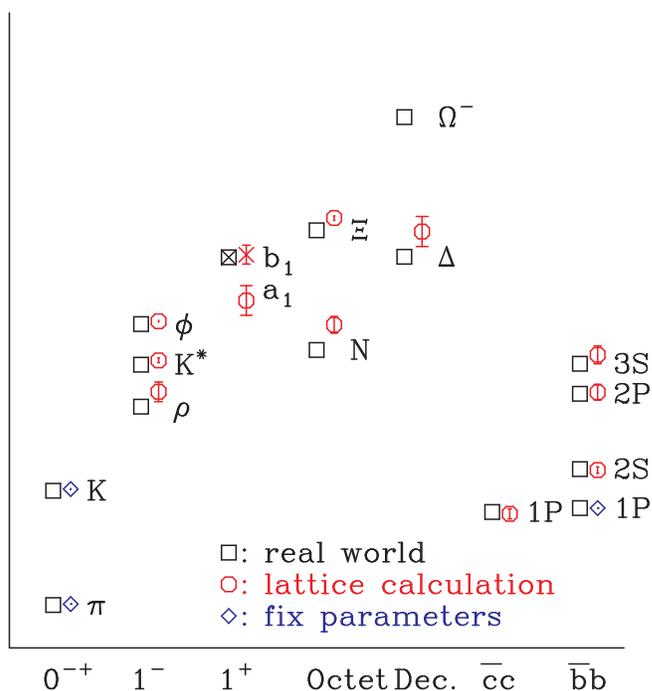}
\end{center}
\caption{
Unquenched hadron masses and splittings compared with experimental values$^{69}$. The $\Upsilon$ and $c\bar c$ columns are differences from the ground state masse as computed by HPQCD and Fermilab groups. The $\pi$, $K$ and $\Upsilon$(1P-1S) masses are used to fix the quark masses and the lattice spacing.\label{theoryvsexperiment2}}
\end{figure}

There are two types of errors in any LQCD computation. Statistical errors and systematic errors. Statistical errors are under control today.
The main source of systematic error are discussed below, they are being addressed by recent computations, and will be removed in the near future.

\begin{itemize}

\item

Discretization. The typical lattice spacing is today of the order of $(2\textrm{GeV})^{-1}$. This introduces discretization errors that are sometime difficult to quantify. One effect, for example, is the breaking of continuous rotational symmetry. One way to reduce discretization effects is by means of Symazik improved actions. 

\item

Quenching. This error is the most difficult to quantify and it is now being eliminated thanks to new algorithms and cheaper computing power. A great deal of effort has been put in this direction in the recent years. Recent computations with dynamical quarks have been a success but the dynamical quark masses are still larger then the physical $u$ and $d$ quark mass. 

\item

Chirality. Both Wilson and Staggered fermions break chiral symmentry and do not allow computations at zero quark mass. Moreover, the more the chiral limit is approached, the more expensive it is to invert the fermionic matrix $Q$. Alternative fermionic discretizations such as domain wall fermions and overlap fermions promise restoration of the chiral simmetry in the continuum limit and provide a better approximation of continuum physics.

\item
Matching. Most experimental quantities are usually expressed in the $\bar{MS}$ scheme, while the lattice computations are performed in a different regularization/renormalization scheme. The procedure to relate one to the other is called matching and it mainly perturbative in nature. When lattice results are converted in the $\bar{MS}$ scheme using one-loop  matching they become affected by matching errors as big as 20\%. The computation of matching beyond one loop is a very challenging task. Some attempts to automate this process via a numerical approach to lattice pertubation theory can be found in \cite{DiRenzo:2004ge} and \cite{Trottier:2003bw}.
It is important to stress that the use of the $\bar{MS}$ scheme is a convention and this matching would not be necessary if the lattice regularization were used everywhere.

\end{itemize}

In LQCD, errors due to contiuum extrapolation, extrapolation to physical masses (both for heavy quarks and light quarks) and finite lattice size are today very much under control and below 2\% for most quantities of phenomenological interest.

Fig.~\ref{theoryvsexperiment} and fig.~\ref{theoryvsexperiment2} show the effect of quenching on the light quark spectrum and how the latest dynamical LQCD computations\cite{Davies2003ik,Aubin:2004wf} agree with experiment.

\section{Conclusion}

The lattice formulation provides a way to regularize, define and compute the Path Integral in a Quantum Field Theory. This formulation and the associated numerical techniques have been of crucial importance in deepening our knowledge of quantum field theories in physical regimes where perturbation theory fails, as in the case of QCD at strong coupling. For example, LQCD computations have played and are still playing an important role in understanding the process of quark confinement\cite{Bernard2000gd,engelhardt}, determining the phase structure of gauge theories\cite{Kogut:2004su}, confronting QCD predictions with experiments\cite{Davies2003ik}, and extracting fundamental physical parameters such as quark masses\cite{Gockeler:2004rp,McNeile:2004cb} and CKM matrix elements\cite{Becirevic:2002zp,Okamoto:2005hq} from experimental results.

Three advantages of LQCD over most analytical techniques is that it is easier to understand, the effect of its approximations can be controlled numerically, and the precision of any computation can be reduced arbitrarily just by increasing the dedicated computing time.

Today most LQCD computations are limited by available computing power. In order to overcome this limitation many groups have been working on designing dedicated machines for LQCD such as APEnext\cite{ape}, QCDOC\cite{qcdoc} and the Earth Simulator\cite{earth}. Other groups have focused on the development of software and algorithms optimized for commodity hardware such as PC clusters\cite{cluster} and building infrastructures for the exchange of field configurations\cite{ildg}.

Most of the algorithms and the examples discussed here and many more are implemented in a free software library called FermiQCD\cite{DiPierro:2000ve,DiPierro:2001yu,DiPierro:2003sz}. Despite its name, FermiQCD is not LQCD specific but it is suitable for generic LQFT computations. It has a modular design and an easy to use syntax very similar to the one we have used in this paper (in fact, algorithms such as the MinRes and the BiCGStab map line by line). Moreover, all of the FermiQCD algorithms are parallel and they can run on distributed memory architectures such as PC clusters. FermiQCD has been used for production grade computations at Fermilab and other institutions.
\vskip 5mm
FermiQCD can be downloaded from {\tt www.fermiqcd.net}.
\vskip 5mm
{\bf Acknowledgements} \\
I wish to acknowledge the Fermilab theory group for many years of fruitful collaboration and specifically thank Paul Mackenzie for his comments about this manuscript. I thank Anthony Duncan, Estia Eichten and the MILC collaboration for giving me permission to reproduce some plots from their papers, and David Kaplan for spotting a typo in my first draft. I am also very grateful to Lim Chee Hok, editor of IJMPA, for asking me to write this review.

\appendix

\section{Useful Formulas in 4D Euclidean Space-Time}

\subsection{Wick rotation}

The Euclidean action is obtained from the Minkowskian one by performing a Wick rotation.
Under this rotation the basic vectors of the theory transform according with the following table
(E for Euclidean, M for Minkowski) 
\index{Wick rotation}
\begin{equation}
\begin{tabular}{|cc|cc|} \hline
E & M & E & M \\ \hline
$x^0$ & $ix^0$ & $x^i$ & $x^i$ \\
$\partial^0$ & $-i\partial _0$ & $\partial^i$ & $\partial _i$ \\
$A^4$ & $-iA_0$ & $A^i$ & $A_i$ \\ 
$F^0i$ & $-iF_{0i}$ & $F^{ij}$ & $F_{ij}$ \\
$\gamma^0 $ & $\gamma ^0$ & $\gamma^i$ & $-i\gamma ^i$ \\
$\gamma^5$ & $\gamma ^5$ & &  \\ \hline
\end{tabular}
\label{wickrot}
\end{equation}
and the integration measure transforms as follow 
\begin{equation}
\exp (-{\cal S}_E)=\exp (i{\cal S}_M)
\end{equation}
where 
\begin{equation}
{\cal S}_E=\int \text{d}^4 x_E {\cal L}_E[...]=-i\int 
\text{d}^4x_M{\cal L}_M[...]
\end{equation}
The choice $\text{d}^4x_E=i\text{d}^4x_M$ can be made, hence 
${\cal L}_E[...]=-{\cal L}_M[...]$

The Euclidean metric tensor is defined as
\begin{equation}
g_E^{\mu \nu }=-\delta ^{\mu \nu }=\text{diag}(-1,-1,-1,-1)
\end{equation}

\subsection{Spin matrices}

\begin{itemize}
\item  Dirac matrices (Dirac representation) \index{Dirac matrices!Euclidean}
\begin{equation}
\begin{tabular}{lll}
$\gamma^0=\left( 
\begin{array}{ll}
{\bf 1} & 0 \\ 
0 & -{\bf 1}
\end{array}
\right)$,\,\, & $\gamma^i=\left( 
\begin{array}{ll}
0 & -i\sigma _i \\ 
i\sigma _i & 0
\end{array}
\right)$,\,\, & $\gamma^5=\left( 
\begin{array}{ll}
0 & 1 \\ 
1 & 0
\end{array}
\right) $ 
\end{tabular}
\end{equation}

\item Dirac matrices (Chiral representation)
\begin{equation}
\begin{tabular}{lll} 
$\gamma^0=\left( 
\begin{array}{ll}
0 & {\bf 1} \\ 
{\bf 1} & 0
\end{array}
\right)$,\,\, & $\gamma^i=\left( 
\begin{array}{ll}
0 & -i\sigma _i \\ 
i\sigma _i & 0
\end{array}
\right)$,\,\, & $\gamma^5=\left( 
\begin{array}{ll}
-{\bf 1} & 0 \\ 
0 & {\bf 1}
\end{array}
\right) $%
\end{tabular}
\end{equation}
All the Euclidean Dirac matrices are hermitian. The following relations hold 
\begin{eqnarray}
g^{\mu \nu } &=&\frac 12\{\gamma^\mu ,\gamma^\nu \} = \delta^{\mu\nu} \\
\sigma^{\mu \nu } &=&\frac i2[\gamma^\mu ,\gamma^\nu ] \\
\gamma^5 &=&\gamma^0\gamma^1\gamma^2\gamma^3
\end{eqnarray}
and all the $\sigma ^{\mu \nu }$ are hermitian.

\item  Projectors 
\begin{equation}
L=\frac{1-\gamma^5}2\qquad R=\frac{1+\gamma^5}2
\end{equation}

\item  Traces 
\begin{eqnarray}
\text{tr}(\gamma^\mu \gamma^\nu ) &=& 4\delta ^{\mu \nu } \\
\text{tr}(\gamma^\mu \gamma^\nu \gamma^\rho ) &=& 0 \\
\text{tr}(\gamma^\mu \gamma^\nu \gamma^\rho 
\gamma^\sigma ) &=& 4(\delta^{\mu \nu }\delta^{\rho
\sigma }-\delta^{\mu \rho }\delta^{\nu \sigma }+\delta^{\mu
\sigma }\delta^{\rho \nu }) \\
\text{tr}(\gamma^5\gamma^\mu\gamma^\nu\gamma^\rho \gamma^\sigma ) 
&=&4\epsilon _E^{\mu \nu \rho\sigma }
\end{eqnarray}
where $\epsilon _E^{0123}=-1.$
\end{itemize}

\end{document}